\documentclass[10pt,aps,prd,floatfix,notitlepage,nofootinbib]{revtex4-1}
\usepackage[utf8]{inputenc}
\usepackage{multirow}
\usepackage{latexsym}
\usepackage{epsfig,graphics,subfig,float}

\usepackage{tabularx,booktabs,caption}% http://ctan.org/pkg/{tabularx,booktabs,caption}

\usepackage{amssymb}
\usepackage{verbatim}
\usepackage{multirow}
\usepackage{color}
\usepackage{tikz}
\usepackage{amsmath,amssymb,amsfonts}

\usepackage{tikz}
\usepackage{bm}
\usetikzlibrary{matrix}

\begin{document}

\title{Generalised nonminimally gravity-matter coupled theory}
\author{Sebastian Bahamonde}
\email{sebastian.beltran.14@ucl.ac.uk}
\affiliation{Department of Mathematics, University College London,
	Gower Street, London, WC1E 6BT, United Kingdom}

\begin{abstract}
In this paper, a new generalised gravity-matter coupled theory of gravity is presented.  This theory is constructed by assuming an action with an arbitrary function $f(T,B,L_m)$ which depends on the scalar torsion $T$, the boundary term $B=\nabla_{\mu}T^{\mu}$ and the matter Lagrangian $L_m$. Since the function depends on $B$ which appears in $R=-T+B$, it is possible to also reproduce curvature-matter coupled models such as $f(R,L_m)$ gravity. Additionally, the full theory also contains some interesting new teleparallel gravity-matter coupled  theories of gravities such as $f(T,L_m)$ or $C_1 T+ f(B,L_m)$. The complete dynamical system for flat FLRW cosmology is presented and for some specific cases of the function, the corresponding cosmological model is studied. When it is necessary, the connection of our theory and the dynamical system of other well-known theories is discussed. 
\end{abstract}

%\pacs{04.30, 04.30.Nk, 04.50.+h, 98.70.Vc}
\date{\today}

\maketitle
%\tableofcontents
	
\section{Introduction}\label{sec:0}
Nowadays, one of the most important challenges in physics is try to understand the current acceleration of the Universe. In 1998, using observations from Supernovae type Ia, it was shown that the Universe is facing an accelerating expansion, changing the way that we understand how our Universe is evolving \cite{Riess:1998cb}. Later, other cosmological observations such as CMB observations \cite{Spergel:2003cb,Spergel:2006hy,Komatsu:2008hk,Komatsu:2010fb}, baryon acoustic oscillations \cite{Eisenstein:2005su} or galaxy clustering \cite{Tegmark:2003ud} also confirmed this behaviour of the Universe. The responsible of this late-time acceleration of the Universe is still not well understood and for that reason it was labelled as the dark energy problem. In general, there are two different approaches which try to deal with this issue. First, one can assume that General Relativity (GR) is always valid at all scales and introduce a new kind of matter which mimics this acceleration. This kind of matter known as ``exotic matter" needs to violate the standard energy conditions to describe the evolution of the Universe. Up to now, this kind of matter has not been discovered in the laboratory. One can say that this approach lies on the idea of changing the right hand side of the Einstein field equations. An alternative approach to understand and study the dark energy is to assume that GR is only valid at certain scales and therefore it needs to be modified. In this approach, the left hand side of the Einstein field equations is modified and there is no need to introduce exotic matter. Different kind of modified theories of gravity have been proposed in the literature to understand the dark energy problem (see the reviews \cite{Nojiri:2006ri,Capozziello:2011et}).\\

One very interesting and alternative theory of gravity is the teleparallel equivalent of general relativity (TEGR) or ``teleparallel gravity". In this theory, the manifold is endorsed with torsion but assumes a zero curvature. The  connection which satisfies this kind of geometry is the so-called ``Weitzenb\"{o}ck" connection, which was first introduced in 1922  \cite{weitzenbock1923invariantentheorie}. It was then showed that this theory is equivalent to GR in the field equations but the geometrical interpretation of gravity is different. In TEGR, there is not geodesic equation as in GR. Instead, forces equations describe the movement of particles under the influence of gravity. Additionally, the dynamical variable is the tetrad instead of the metric as in GR. For more details about TEGR, see \cite{Hayashi:1977jd,deAndrade:1997qt,deAndrade:2000kr,Arcos:2005ec,Obukhov:2002tm,Maluf:2013gaa} and also the book~\cite{AP}. Similarly as in GR, there are also modified theories starting from the teleparallel approach. The most famous teleparallel modified theory is $f(T)$ gravity (where $T$ is the scalar torsion) which can describe very well the current acceleration of the Universe and also other cosmological observations (see~\cite{Bengochea:2008gz,Ferraro:2006jd,Bengochea:2010sg,Li:2011wu,Wu:2010av,Wu:2010mn,Dent:2011zz,Chen:2010va,Cai:2011tc,Bamba:2010wb,Biswas:2015cva,Wu:2010xk,Nunes:2016qyp,Zubair:2014xsa,Hohmann:2017jao} and also the review \cite{Cai:2015emx}). The TEGR action contains the term $T$ so $f(T)$ gravity is a straightforward generalisation of it. This theory is analogous to the well-known $f(R)$ gravity, where instead of having the scalar curvature $R$ in the action, a more general theory with an arbitrary function which depends on $R$ is introduced. These two theories are analogous but mathematically they are very different. As we pointed out before, the TEGR field equations are equivalent to the Einstein field equations. However, their generalisations $f(R)$ and $f(T)$ gravity have different field equations. Further, $f(R)$ gravity is a 4th order theory and $f(T)$ gravity is a 2nd order theory. This characteristic can be understood using the fact that $R=-T+B$, where $B$ is a boundary term. Hence, a linear combination of $R$ or $T$ in the action will produce the same field equations since $B$ will not contribute to it. However, when one modifies the action as an arbitrary function $f(T)$ or $f(R)$, there will be a difference in their field equations due to the fact that now the boundary term $B$ contributes. This was fully studied in \cite{Bahamonde:2015zma} where the authors introduced a new theory, the so-called $f(T,B)$ gravity, which can recover either $f(T)$ gravity or $f(T,B)=f(-T+B)=f(R)$ as special cases. Flat FLRW cosmology of this theory was studied in \cite{Bahamonde:2016cul,Bahamonde:2016grb}. \\

Other kinds of modified theories of gravity have been considered in the literature. Some interesting ones are theories with non-minimally coupling between matter and gravity. In standard metric approach, some alternatives models have been proposed such as $f(R,\mathcal{T})$ \cite{Harko:2011kv}, where $\mathcal{T}$ is the trace of the energy-momentum tensor or non-minimally coupled theories between the curvature scalar and the matter Lagrangian $f_1(R)+f_2(R)L_m$ \cite{Bertolami:2007gv}. Further, another more general theory  is the so-called $f(R,L_m)$ where now an arbitrary function of $R$ and $L_m$ is considered in the action \cite{Harko:2011kv}. Along the lines of those theories, modified teleparallel theories of gravity where couplings between matter and the torsion scalar have been also considered. Some important theories are for example: $f(T,\mathcal{T})$ gravity \cite{Harko:2014aja} and also non-minimally couplings between the torsion scalar and the matter Lagrangian theory $f_1(T)+f_2(T)L_m$ \cite{Harko:2014sja}. Along this line, in this paper, we present a new modified teleparallel theory of gravity based on an arbitrary function $f(T,B,L_m)$ where $L_m$ is the matter Lagrangian. In this theory, we have the possibility of for example recover $f(-T+B,L_m)=f(R,L_m)$ or a new generalisation of \cite{Harko:2014sja} in the teleparallel framework with a function $f(T,L_m)$ depending on $T$ and $L_m$. The later new theory is the analogous theory as $f(R,L_m)$ gravity. We will explicitly  discuss about how those models are related, with $B$ being the main ingredient which connects both the metric and tetrad approaches. \\

After formulating the new $f(T,B,L_m)$ theory, the conservation equation is obtained and exactly as in $f(R,L_m)$, the conservation equation in $f(T,B,L_m)$ theory is not always valid. It will be proved that for the flat FLRW case and assuming $L_m=-2\rho$, the conservation equation is conserved exactly as happens in $f(R,L_m)$ or in $f_1(R)+f_2(R)L_m$ (see \cite{Bertolami:2008ab,Azevedo:2016ehy}). The main aim of this paper is to also formulate the dynamical system of this new generalise theory, which is in general a 10-dimensional one. This dynamical system is a generalisation of different models such as the ones studied in \cite{Azevedo:2016ehy,An:2015mvw,Carloni:2015lsa}. After formulating the full dynamical system, different special cases are recovered. Some of them have been studied in the past, hence we only mention how our dimensionless variables are related to them and then we show that our dynamical system becomes them for the special case studied. Then, using  dynamical system techniques, we will study new cases that can be constructed from our action. Similarly as in $f(R,L_m)$ (see \cite{Azevedo:2016ehy}), a power-law and a exponential kind of coupling between $L_m$ and $T$ is studied. Additionally, another new kind of couplings between the boundary term $B$ and $L_m$ are studied. For this theory, we study different power-law models with  $f(T,B,L_m)=C_1T+C_5B^{s}+(C_4+C_4 B^{q})L_m$. This model depends highly on the power-law parameters $s$ and $q$. The critical points and their stability are then studied for different models. For the readers interested on dynamical systems in cosmology, see the review \cite{Bahamonde:2017ize} and also see \cite{Bahamonde:2018miw,Bahamonde:2015hza} for further applications to dynamical systems in modified teleparallel models with the boundary term $B$.
\\

The notation of this paper is the following: the natural units are used so that $\kappa=1$ and the signature of the metric is $\eta_{ab}=(+1,-1,-1,-1)$. The tetrad and the inverse of the tetrad are labelled as $e^{a}_{\mu}$ and $E_a^{\mu}$ respectively where Latin and Greek indices represent tangent space and space-time coordinates respectively. \\

This paper is organized as follows: Sec.~\ref{sec:1} is devoted to present a very brief review of teleparallel theories of gravity and some interesting modified theories than can be constructed from this approach. In Sec.~\ref{sec:2} is presented the new generalised gravity-matter coupled theory of gravity known as $f(T,B,L_m)$ where $T,B$ and $L_m$ are the scalar torsion, the boundary term and the matter Lagrangian respectively. The corresponding field equations of the theory and the flat FLRW cosmological equations are also derived in this section. In Sec.~\ref{sec:3} is presented the dynamical system of the full model and for some specific theories, the corresponding dynamical analysis of them is performed. Finally, Sec.~\ref{sec:4} concludes the main results of this paper.

	\section{Teleparallel gravity and its modifications}\label{sec:1}
Let us briefly introduce the teleparallel equivalent of general relativity (TEGR) and some important modifications under this theory. Basically, this theory is based on the idea of having a globally flat manifold (zero curvature) but with a non-trivial geometry for having a non-zero torsion tensor. Hence, the concept of paralellism is globally defined in TEGR. The dynamical variable of this theory is the tetrad which defines orthonormal vectors at each point of the manifold and they are directly related with the metric as follows
\begin{align}
g_{\mu\nu} = \eta_{ab} e^a{}_{\mu} e^b{}_{\nu} \,,
\label{met}
\end{align}
where $\eta_{ab}$ is the Minkowski metric. The connection which defines a globally flat curvature with a non-vanishing torsion is the so-called Weitzenb\"{o}ck connection $W_{\mu}{}^{a}{}_{\nu}$, which defines the torsion tensor as taking its anti-symmetric part, namely
\begin{align}
T^{a}{}_{\mu\nu} &= W_{\mu}{}^{a}{}_{\nu} - W_{\nu}{}^{a}{}_{\mu} =
\partial_{\mu} e_{\nu}^{a} - \partial_{\nu}e_{\mu}^{a} \,.
\label{eq:tor}
\end{align}
Let us clarify here that the above definition is not the most general form of the torsion tensor. The most general definition also contains the spin-connection which needs to be pure gauge in order to fulfil the condition of teleparallelism (zero curvature). In this paper is assumed that the spin-connection is identically zero.\\
The TEGR action is defined with the so-called torsion scalar $T$ as follows
\begin{align}\label{action1}
S_{\rm TEGR} = \int e\,T \, d^4x+ S_{\rm m} \,,
\end{align}
where $e=\textrm{det}(e_{\mu}^{a})$ and $S_{\rm m}$ is the matter action. The torsion scalar is defined as the contraction of the super-potential 
\begin{align}
S^{abc} = \frac{1}{4}(T^{abc}-T^{bac}-T^{cab})+\frac{1}{2}(\eta^{ac}T^b-\eta^{ab}T^c) 
\end{align}
with the torsion tensor as $T=T_{abc}S^{abc}$. Here, $T_{\mu}=T^{\lambda}{}_{\lambda\mu}$ is the so-called torsion vector. The definition of $T$ comes directly from the condition of zero-curvature where one arrives that the Ricci scalar is directly linked with it via
\begin{eqnarray}
R=-S^{abc}T_{abc}+\frac{2}{e}\partial_{\mu}(eT^{\mu})=-T+B\,,\label{RTB}
\end{eqnarray}
where $B$ refers to the boundary term which connects the Ricci scalar with the torsion scalar. From (\ref{action1}) and the above relationship, one can directly notice that the TEGR is equivalent to the Einstein-Hilbert action up to a boundary term. Hence, TEGR is an alternative formulation of gravity which reproduces the same field equations as GR. Although, the geometrical interpretation of these theories are different. GR lies in a manifold with a non-zero curvature (in general) with a zero torsion tensor whereas TEGR is the opposite. Moreover, geodesic equations are replaced by forces equations in TEGR (see \cite{AP} for more details about this theory). \\
 A straightforward generalisation of the action (\ref{action1}) is to replace $T$ by an arbitrary function of $f$ which depends on $T$, namely
 \begin{align}\label{action22}
 S_{f(T)} = \int e\,f(T) \, d^4x+ S_{\rm m} \,.
 \end{align}
The former theory is the most popular modification of TEGR and it was firstly introduced in \cite{Ferraro:2006jd} with the aim to study inflation in cosmology. In some sense, this generalisation is analogous as the famous modification of GR, the so-called $f(R)$ gravity, where instead of having $R$ in the Einstein-Hilbert action, an arbitrary function of $R$ is introduced in the action. The formulation described here for $f(T)$ gravity where the spin-connection is identically zero is not invariant under Lorentz transformations. This is due to the fact that $T$ itself is not invariant under local Lorentz transformations so $f(T)$ gravity will also have this property \cite{Sotiriou:2010mv,Li:2010cg}. In standard TEGR where $T$ is in the action, this problem is not important since the action only differs by a boundary term with respect to the Einstein-Hilbert action so one can say that this theory is quasi-invariant under local Lorentz transformation. The problem of the loose of the Lorentz invariant produces that two different tetrads could give rise different field equations so it depends on the frame used. For example, the flat FLRW in spherical coordinates give rise to different field equations as in Cartesian coordinates. At the level of the field equations, this problem can be alleviated by choosing ``good tetrads" as it was introduced in \cite{Tamanini:2012hg}. In this approach, one needs to rotate the tetrad fields and fix it accordingly depending on the geometry studied. In~\cite{Krssak:2015oua}, it was proposed a new approach of teleparallel theories of gravity where a non-zero spin-connection is assumed giving rise to a covariant version of $f(T)$ gravity. Both approaches should arrive at the same field equations and since almost all the works based on $f(T)$ gravity used the approach presented above, we will continue using this approach.  

It is also possible to create other kind of modifications of teleparallel theories of gravity. A very interesting modification theory is given by the following action \cite{Bahamonde:2015zma} 
 \begin{align}\label{action222}
S_{f(T,B)} = \int e\,f(T,B) \, d^4x+ S_{\rm m} \,,
\end{align}
where now the function also depends on the boundary term $B$. Under this theory, it is possible to recover either $f(-T+B)=f(R)$ gravity or $f(T)$ gravity. Moreover, the theory $f(T,B)=C_1T+f_1(B)$ can also be obtained from this action. From this theory one can directly see how $f(R)$ and $f(T)$ are connected by this boundary term. Since $R=-T+B$, only if a linear combination of $R$ and $T$ is 
assumed in the action (TEGR or GR), we will have equivalent theories at the level of the field equations. It is known that $f(R)$ gravity is a 4th order theory whereas $f(T)$  gravity is a 2nd order theory. Hence, $f(T,B)$ gravity is also a 4th order theory. $f(T)$ and $f(R)$ gravity have different field equation orders since the difference comes from integrating by parts twice the boundary term $B$.

\section{$f(T,B,L_m)$ gravity}\label{sec:2}
\subsection{General equations}
 Inspired by the theories described in \cite{Harko:2010mv} in the curvature approach and also from $f(T,B)$ gravity, let us now consider the following gravity model
    	\begin{eqnarray}
    	S_{f(T,B,L_m)}=\int e f(T,B,L_m)\,d^{4}x\,,\label{action} 
    	\end{eqnarray}
where the function $f$  depends on the scalar curvature $T$, the boundary term $B$ and the matter Lagrangian $L_m$. The energy-momentum tensor of matter $\mathcal{T}^{\beta}_{a}$ is defined as
\begin{equation}
\mathcal{T}^{\beta}_{a}=-\frac{1}{2e} \frac{\delta (eL_{m})}{\delta e^{a}_{\beta}}\,.
\end{equation}  
Now, we will assume that the matter Lagrangian depends only on the components of the tetrad (or metric) and not on its derivatives, giving us
\begin{eqnarray}
2\mathcal{T}^{\beta}_{a}=-L_{m}E_{a}^{\beta}-\frac{\partial L_m}{\partial e_{\beta}^{a}}\,.\label{energym}
\end{eqnarray}    
 Now, by a variation of action (\ref{action}) with respect to the tetrad, we obtain 
	\begin{eqnarray}
	\delta S_{f(T,B,L_m)}&=&\int  \Big[ef_{T}\delta T+ef_{B}\delta B+ef_{L}\frac{\delta  L_m}{\delta e_{\beta}^{a}}\delta e_{\beta}^{a}+f \delta e\Big]d^{4}x\,,\\
&=&\int e \Big[f_{T}\delta T+f_{B}\delta B-f_{L}\Big(2\mathcal{T}_{a}^{\beta}+L_{m}E_{a}^{\beta}\Big)\delta e_{\beta}^{a}+fE_{a}^{\beta}\delta e_{\beta}^{a}\Big]d^{4}x\,,	\label{action2} 
	\end{eqnarray}
where we have used Eq.~(\ref{energym}) and $f_T=\partial f/\partial T,f_B=\partial f/\partial B$ and $f_L=\partial f/\partial L_{m}$. Variations with respect to the torsion scalar and the boundary term are given by \cite{Bahamonde:2015zma}
\begin{eqnarray}
	e f_{T} \delta T &=&-4e\Big[
\frac{1}{e}\partial_{\mu}( eS_{a}\,^{\mu\beta})f_{T}-f_{T}T^{\sigma}\,_{\mu a}S_{\sigma}\,^{\beta\mu}+	(\partial_{\mu}f_{T})S_{a}\,^{\mu\beta}
	\Big]\delta e^{a}_{\beta} \,,\label{deltaT}\\
	ef_{B}\delta B&=& e\Big[2E_{a}^{\sigma}\nabla^{\beta}\nabla_{\sigma}f_{B}-2E_{a}^{\beta}\Box f_{B}-Bf_{B}E_{a}^{\beta}-4(\partial_{\mu}f_{B})S_{a}\,^{\mu\beta}\Big]\delta e_{\beta}^{a} \,,
	\label{deltaB}
\end{eqnarray}
so that by imposing $\delta S_{f(T,B,L_m)}=0$, we obtain the $f(T,B,L_m)$ field equations given by
\begin{eqnarray}
2E_{a}^{\sigma }\nabla ^{\beta}\nabla
_{\sigma }f_{B}-2E_{a}^{\beta }\Box f_{B}-Bf_{B}E_{a}^{\beta }-4\Big[(\partial_{\mu}f_{T})+(\partial_{\mu}f_{B})\Big]S_{a}\,^{\mu\beta}-4f_{T}\Big(e^{-1}\partial_{\mu}(e S_{a}\,^{\mu\beta})-T^{\sigma}\,_{\mu a}S_{\sigma}\,^{\beta\mu}\Big)\nonumber\\
+f E_{a}^{\beta}-f_{L}L_{m}E_{a}^{\beta}=2 f_{L}\mathcal{T}_{a}^{\beta}\,.\label{fieldeq1}
\end{eqnarray}
The above field equations can be also written only in space-time indices by contracting it by $e^{a}_{\lambda}$ giving us
\begin{eqnarray}
2\nabla ^{\beta}\nabla
_{\lambda }f_{B}-2\delta_{\lambda}^{\beta}\Box f_{B}-Bf_{B}\delta_{\lambda}^{\beta}-4\Big[(\partial_{\mu}f_{T})+(\partial_{\mu}f_{B})\Big]S_{\lambda}\,^{\mu\beta}-4f_{T}e^{a}_{\lambda}\Big(e^{-1}\partial_{\mu}(e S_{a}\,^{\mu\beta})-T^{\sigma}\,_{\mu a}S_{\sigma}\,^{\beta\mu}\Big)\nonumber\\
+f \delta^{\beta}_{\lambda}-f_{L}L_{m}\delta^{\beta}_{\lambda}=2 f_{L}\mathcal{T}_{\lambda}^{\beta}\,.\label{fieldeq2}
\end{eqnarray}
From these field equations, one can directly recover teleparallel gravity by choosing $f(T,L_m)=T+L_m$ which gives us the same action as (\ref{action1}). Moreover if we choose $f(T,L_m)=T+f_1(T)+ (1+\lambda f_2(T))L_m$ we recover the non-minimal torsion-matter coupling extension of $f(T)$ gravity presented in \cite{Harko:2014sja}. Note that in our case, we have assumed that the matter Lagrangian does not depend on the derivatives of the tetrads, which according to \cite{Harko:2014sja} is equivalent as having
\begin{equation}
\frac{\partial L_m}{\partial (\partial_{\mu}e_{\rho}^a)}=0\,.
\end{equation}
Let us now study the conservation equation for this theory. First, we will use that $R^{\beta}_{\lambda}=G_{\lambda}^{\beta}+\frac{1}{2}(B-T)\delta_{\lambda}^{\beta} $, where $G_{\lambda}^{\beta}$ is the Einstein tensor. Using this relationship, we can rewrite the field equation (\ref{fieldeq2}) as follows
\begin{equation}
H_{\lambda\beta}:=f_T G_{\lambda \beta}+\nabla_{\lambda}\nabla_{\beta}f_{B}-g_{\lambda\beta}\square f_{B}-\frac{1}{2}\Big(Tf_{T}+Bf_B+L_m f_L-f\Big)g_{\lambda\beta}-2X_{\nu}S_{\lambda}{}^{\nu}{}_{\beta}=f_{L}\mathcal{T}_{\lambda\beta}\,,\label{eqq}
\end{equation}
where for simplicity we have also introduced the quantity
\begin{eqnarray}
X_{\nu}=(f_{BT}+f_{BB}+f_{BT})\nabla_{\nu}B+(f_{TT}+f_{TB}+f_{TL})\nabla_{\nu}T+(f_{TL}+f_{BL}+f_{LL})\nabla_{\nu}L_{m}\,.
\end{eqnarray}
By taking covariant derivative of $H_{\lambda\beta}$ and after some simplifications, we find that
\begin{eqnarray}
\nabla^{\lambda}H_{\lambda\beta}=2S^{\sigma\rho}{}_{\lambda}K_{\beta\sigma\rho}X^{\lambda}-\frac{1}{2}g_{\lambda\beta}\nabla^{\lambda}(L_mf_L)=-\frac{1}{2}g_{\lambda\beta}\nabla^{\lambda}(L_mf_L)\,,
\end{eqnarray}
where we have used the fact that the energy-momentum tensor is symmetric and hence $S^{\sigma\rho}{}_{\lambda}K_{\beta\sigma\rho}X^{\lambda}=0$. The latter comes from the fact that field equations are symmetric, and hence the energy-momentum tensor is also symmetric. Now, we will find the condition that $f$  needs to satisfy in order to have the standard conservation equation for the energy momentum tensor, i.e., $\nabla_\mu \mathcal{T}^{\mu\nu}=0$. By taking covariant derivative in \eqref{eqq} and assuming $\nabla_\mu\mathcal{T}^{\mu\nu}=0$, one gets that the standard conservation equation for the energy-momentum tensor is satisfied if the function $f$ satisfy the following form
\begin{eqnarray}
\Big(2\mathcal{T}_{\mu\nu}+g_{\mu\nu}L_m \Big)\nabla^{\mu}f_L=-e^{a}_{\mu}g_{\beta\nu}\frac{\partial L_m}{\partial e^{a}_{\beta}}\nabla^{\mu}f_L=0\,,\label{cons}
\end{eqnarray}
which matches with the conservation equation presented in \cite{Harko:2010mv}. Note that in our case, we have defined the energy-momentum tensor in a different way so that there is a minus sign of difference between Eq.~(13) presented in \cite{Harko:2010mv} and the above equation. Thus, in general, $f(T,B,L_m)$ is not covariantly conserved and depending on the metric, the model and the energy-momentum tensor, this theory may or may not be conserved.
Hereafter, we will consider that the matter is described by a perfect fluid whose energy-momentum tensor is given by
\begin{eqnarray}
\mathcal{T}_{\mu\nu}&=&(\rho+p)u_{\mu}u_{\nu}-pg_{\mu\nu}\,.\label{Tmunu}
\end{eqnarray}
Here, $\rho$ and $p$ are the energy density and the pressure of the fluid respectively and $u_{\mu}$ is the 4-velocity measured by a co-moving observer with the expansion so that it satisfies $u_{\mu}u^{\mu}=1$. For a perfect fluid, if one assumes that in the proper frame where the particle is static, the matter Lagrangian is invariant under arbitrary rescaling of time coordinate \cite{Avelino:2018qgt}. Therefore, from \eqref{energym}, one gets $T_{00}=\rho=-(1/2)L_m$ which is equivalent as having $L_m=-2\rho$. This is a ``natural choice" for a perfect fluid (see \cite{Harko:2014sja,Bertolami:2008ab,Avelino:2018qgt} for more details).  Hence, from Eq. (\ref{cons}) we can directly conclude that the conservation law will be always satisfied when flat FLRW and a perfect fluid are chosen without depending on the model for the function $f(T,B,L_m)$. This statement was also mentioned in \cite{Harko:2010mv}, which is a special case of our theory, explicitly when $f(T,B,L_m)=f(-T+B,L_m)=f(R,L_m)$. 
\subsection{Flat FLRW cosmology}
In this section we will briefly find the corresponding  modified flat FLRW cosmology of our theory. Consider a spatially flat FLRW cosmology whose metric is represented by
\begin{eqnarray}
ds^2&=&dt^2-a(t)^2(dx^2+dy^2+dz^2)\,,\label{FRW}
\end{eqnarray}
where $a(t)$ is the scale factor of the universe. The tetrad corresponding to this space-time in Cartesian coordinates reads
\begin{eqnarray}
e_{\beta}^{a}=\textrm{diag}(1,a(t),a(t),a(t))\,.\label{FRWtetrad}
\end{eqnarray}
For the space-time given by (\ref{FRW}), the modified FLRW equations become
\begin{eqnarray}
3H^2 (3f_{B}+2 f_{T})-3 H \dot{f}_{B}+3 f_{B}\dot{H}+\frac{1}{2} f&=&0\,,\label{eq1a}\\
(3 f_{B}+2f_{T}) (3H^2+\dot{H})+2 H \dot{f}_{T}-\ddot{f}_{B}+\frac{1}{2} f&=&-f_{L} (p+\rho) \,,\label{eq2}
\end{eqnarray}
where $H=\dot{a}/a$ is the Hubble parameter and dots represent derivation with respect to the cosmic time. Note that the terms $\dot{f}_{B}=f_{BB}\dot{B}+f_{BT}\dot{T}+f_{BL}\dot{L}_m$ and $\dot{f}_{T}=f_{BT}\dot{B}+f_{TT}\dot{T}+f_{TL}\dot{L}_m$. It is clear that when $f(T,B,L_m)=T+L_m=T-2\rho$, one recovers standard TEGR (or GR) plus matter. The energy density of matter does not appear in \eqref{eq1a} explicitly since it is implicitly considered in the term $f/2$. When $f(T,B,L_m)=f(-T+B,L_m)=f(R,L_m)$, the above equations are the same as the ones reported in \cite{Azevedo:2016ehy}. Note that in the latter paper, the authors used another signature notation $\eta_{ab}=(-+++)$, so that one needs to change $R\rightarrow -R$ to match those equations. \\
As a consequence of the conservation law holds when considering a perfect fluid as a matter content of the universe, we also know that the standard continuity equation is valid in our case. Hence, we have that the fluid satisfies
\begin{eqnarray}
\dot{\rho}+3H(\rho+p)=0\,.\label{conserv}
\end{eqnarray} 
Let us now assume a barotropic equation of state $p=w \rho$, so that we can directly find that the energy density of the fluid behaves as
\begin{eqnarray}
\rho(t)&=&\rho_0 a(t)^{-3(1+w)}\,,\label{rho}
\end{eqnarray}
where $\rho_0$ is an integration constant. It is also useful to note that the scalar torsion and the boundary term in this space-time satisfy the relationship (\ref{RTB}), namely
\begin{eqnarray}
T=-6H^2\,,\quad B=-6 (\dot{H}+3 H^2)\,, \rightarrow R=-T+B=-6 (\dot{H}+2 H^2)\,.\label{TBR}
\end{eqnarray}

\section{Dynamical systems}\label{sec:3}
\subsection{Dynamical system for the full theory}
In this section we will explore the dynamical system of different theories of gravity coupled with matter. To do this, we will first study the dynamical system of the general modified FLRW by using the conservation equation given by (\ref{conserv}) and also the first modified FLRW equation (\ref{eq1a}). By replacing the boundary term given by Eq.~(\ref{TBR}) in (\ref{eq1a}) and expanding the derivatives of $f$ we get
\begin{eqnarray}
6H^2f_{T}-3H\Big(f_{BB}\dot{B}+f_{BT}\dot{T}+6H(1+w)\rho f_{BL}\Big)-\frac{1}{2}Bf_{B}+\frac{1}{2}f=0\,.\label{eq1}
\end{eqnarray}
where we have used the conservation equation (\ref{conserv}) to replace $\dot{L}_{m}=-2\dot{\rho}=6H\rho(1+w)$. Let us now introduce the following dimensionless variables
\begin{eqnarray}
x_1=\frac{\dot{T} f_{TB}}{2 H f_{T}}\,,\quad x_2=\frac{\dot{B} f_{BB}}{2 H f_{T}}\,,\quad x_3=\frac{\dot{B}f_{BT}}{2Hf_{T}}\,,\quad x_{4}=\frac{\dot{T} f_{TT}}{2 H f_{T}}\,,\quad y_1=\frac{B f_{B}}{12 H^2 f_{T}}\,,\quad y_2=\frac{Tf_B}{12 H^2 f_{T}}=-\frac{f_B}{2 f_{T}}\,,\\
z=-\frac{f}{12 f_{T} H^2}\,,\quad \phi =\frac{3 (w+1) \rho f_{BL}}{f_{T}}\,,  \quad  \alpha=\frac{3 (w+1) \rho f_{TL}}{f_{T}}\,,\quad \theta=\frac{(w+1)  \rho f_{L}}{2 f_{T}H^2}\,,\label{variables}
\end{eqnarray} 
These dimensionless variables were chosen with the aim of having a similar variables as the ones presented in \cite{Azevedo:2016ehy}. Further, using these variables will help us to compare both theories in the limit case where $f(T,B,L_m)= f(-T+B,L_m)=f(R,L_m)$. Using these variables, the Friedmann constraint given by (\ref{eq1}) becomes
\begin{equation}
x_1+x_2+y_1+z+\phi=1\,.\label{Fcons}
\end{equation}
Moreover, using the dimensionless variables, we can find the following useful relations
\begin{eqnarray}
\frac{\dot{f}_{T}}{2 Hf_{T}}&=&x_3+x_4+\alpha\,,\label{id1}\\
\frac{\dot{f}_{B}}{2Hf_{T}}&=& x_1+x_2+\phi\,,\label{id2}\\
\frac{\ddot{f}_{B}}{2 H^2f_{T}}&=&(x_1+x_2+\phi) \Big[\frac{y_1}{y_2}-3+2 (\alpha +x_3+x_4)\Big]+\frac{dx_1}{dN}+\frac{dx_2}{dN}+\frac{d\phi}{dN}\,,\label{id3}\\
\frac{\dot{H}}{H^2}&=&\frac{y_1}{y_2}-3\,,\label{id4}
\end{eqnarray}
where we have defined $N=\ln a$ as the number of e-folding so that $d/dt=H d/dN$.
The effective state matter and the deceleration parameter can be written in terms of these dimensionless parameters as follows
\begin{eqnarray}
w_{\rm eff}&=&\frac{p_{\rm total}}{\rho_{\rm total}}=-\Big(\frac{2\dot{H}}{3H^2}+1\Big)=1-\frac{2}{3}\frac{ y_1}{ y_2}\,,\label{weff}\\
		\tilde{q}&=&-\frac{\dot{H}}{H^2}-1=2-\frac{y_1}{y_2}\,.
\end{eqnarray}
For acceleration universes, one needs that $\tilde{q}<0$ or equivalently $w_{\rm eff}<-1/3$.

 By replacing the identities (\ref{id1})-(\ref{id4}) and the dimensionless variables defined as (\ref{variables}) into the second modified Friedmann equation (\ref{eq2}), we get 
\begin{equation}
\frac{dx_1}{dN}+\frac{dx_2}{dN}+\frac{d\phi}{dN}=-(x_1+x_2+\phi) \left(2( \alpha+x_3+x_4)+\frac{y_1}{y_2}-3\right)+2 (\alpha+x_3+x_4)+\frac{y_1}{y_2}+3 y_1-3 z+\theta\,.
\end{equation}
The conservation equation (\ref{conserv}) can be also written in terms of the dimensionless variables, which yields
\begin{equation}
\frac{d\phi}{dN}=\phi\Big[ 3 (w+1) (2 \beta_{BLL}-1)+2x_3(\beta_{BBL}-1)+2x_4( \beta_{TBL}-1)-2 \alpha\Big]\,,
\end{equation}
where we have defined
\begin{eqnarray}
\beta_{LL}= \frac{\rho f_{LL}}{f_{L}}\,,\quad \beta_{BLL}=\frac{\rho  f_{BLL}}{f_{BL}}\,,\quad \beta_{BBL}=\frac{f_{T}f_{BBL}}{f_{BL}f_{TB}}\,,\quad \beta_{TBL}=\frac{f_{T}f_{TBL}}{f_{BL}f_{TT}}\,, \\
 \beta_{TLL}= \frac{f_{T}f_{TLL}}{f_{BL}^2}\,,\quad \beta_{TTL}= \frac{f_{T}f_{TTL}}{f_{BL}f_{TB}}\,, \quad \alpha_{TB}= \frac{Bf_{TB}}{2f_{T}}\,,\quad \alpha_{TTT}= \frac{f_{T}f_{TTT}}{f_{TT}^2}\,,\\
\quad \alpha_{BBB}= \frac{f_{T}f_{BBB}}{f_{BB}f_{TB}}\,,\quad \alpha_{TTB}= \frac{f_{T}f_{TTB}}{f_{TB}^2}\,,\quad\alpha_{TBB}= \frac{f_{T}f_{TBB}}{f_{TB}^2}\,,\quad \alpha_{TBL}= \frac{f_{T}f_{TBL}}{f_{BL}f_{TB}}\,.
\end{eqnarray}
The other quantities defined above will be useful in the full dynamical system equations. Using the dimensionless variables, the identities mentioned before and the above definitions, one can find the dynamical system. This procedure is very involved since the dynamical system is a 10 dimensional one. After all of those computations, the system can be summarized with the following equations:
\begin{eqnarray}
\frac{dx_1}{dN}&=&2 \left(\alpha_{TTB} x_{1}^2+\alpha_{TBB} x_{1} x_{3}+x_{3}\right)+y_{1} \left(\frac{4 \alpha_{TB}}{y_{2}}-\frac{2 x_{1}}{y_{2}}\right)-4 \alpha_{TB} (\alpha+x_{3}+x_{4}+6)+\frac{12 \alpha_{TB} y_{2} (\alpha+x_{3}+x_{4}+3)}{y_{1}}\nonumber\\
&&+2 \beta_{TBL} x_{4} \phi\,,\label{x1'}\\
\frac{dx_2}{dN}&=&\theta+\phi \left(3 (w+2)-6 (w+1) \beta_{BLL}-2 \beta_{BBL} x_{3}-4 \beta_{TBL} x_{4}-\frac{y_{1}}{y_{2}}\right)-2 \alpha_{TTB} x_{1}^2+\alpha (4 \alpha_{TB}-2 x_{1}-2 x_{2}+2)\nonumber\\
&&+y_{1} \left(\frac{x_{1}}{y_{2}}-\frac{x_{2}}{y_{2}}-\frac{4 \alpha_{TB}}{y_{2}}+\frac{1}{y_{2}}-3\right)+x_{1} (-2 (\alpha_{TBB}+1) x_{3}-2 x_{4}+3)+x_{2} (-2 x_{3}-2 x_{4}+3)\nonumber\\
&&+4 \alpha_{TB} (x_{3}+x_{4}+6)-\frac{12\alpha_{TB}y_2 (\alpha+x_3+x_4+3)}{y_1}+2 x_{4}-3 z\,,\label{x2'}\\
\frac{dx_3}{dN}&=&-\frac{x_3}{x_2} \Big[-2 \alpha-\theta+6 (w+1) \beta_{BLL} \phi-3 (w+2) \phi+2 \alpha_{TTB} x_{1}^2-\frac{2 \alpha_{TBB} x_{1}^2 y_{1} \left(y_{1} \left(\frac{x_{1}}{\alpha_{TB}}-2\right)+y_{2} \left(6-\frac{x_{3}}{\alpha_{TB}}\right)\right)}{y_{2} (y_{1}-3 y_{2})}\nonumber\\
&&+4 \beta_{BBL} x_{3} \phi-2 x_{3}+4 \beta_{TBL} x_{4} \phi-2 x_{4}+\frac{y_{1} (3 y_{2}+\phi-1)}{y_{2}}+3 z\Big]+\frac{x_{1} x_{3} \left(x_{1} y_{1}^2-x_{3} y_{1} y_{2}-3 \alpha_{TB} (y_{1}-3 y_{2})^2\right)}{\alpha_{TB} x_{2} y_{2} (y_{1}-3 y_{2})}\nonumber\\
&&+2 \alpha_{TTB} x_{1} x_{3}-\frac{2 \alpha_{BBB} x_{3} (-x_{1} y_{1}+x_{3} y_{2}+2 \alpha_{TB} (y_{1}-3 y_{2}))}{y_{2}}-2 \alpha x_{3}+2 (\alpha_{TBB}-1) x_{3}^2\nonumber\\
&&+2 \alpha_{TBL} x_{3} \phi-2 x_{3} x_{4}-\frac{x_{3} y_{1}}{y_{2}}+3 x_{3}\,,\label{x3'}\\
\frac{dx_4}{dN}&=&2 \beta_{TTL}x_{1} \phi+2 \alpha_{TTB} x_{1} x_{3}+x_{4} \left(\frac{y_{1} (x_{3} y_{2}-x_{1} y_{1})}{\alpha_{TB} y_{2} (y_{1}-3 y_{2})}-2 x_{3}+\frac{2 y_{1}}{y_{2}}-6\right)-2 \alpha x_{4}+2 (\alpha_{TTT}-1) x_{4}^2\,,\label{x4'}
\end{eqnarray}
\begin{eqnarray}
\frac{dy_1}{dN}&=&-\frac{y_1 (x_1+x_2+2 y_1+\phi )}{y_2}-2 y_1 (\alpha +x_3+x_4-3)+\frac{x_{3} y_{1}}{\alpha_{TB}}\,,\label{y1'}\\
\frac{dy_2}{dN}&=&\frac{x_{1} y_{1}}{\alpha_{TB}}-x_{1}-x_{2}-2 (y_{2} (\alpha+x_{3}+x_{4}-3)+y_{1})-\phi\,,\label{y2'}\\
\frac{dz}{dN}&=&-\theta+\frac{x_{1} y_{1}}{2 \alpha_{TB} y_{2}}-2 z (\alpha+x_3+x_4-3)-\frac{2 y_1 z}{y_2}-\frac{x_{3} y_{1}}{\alpha_{TB}}\,,\label{z'}\\
\frac{d\phi}{dN}&=&\phi \Big[3 (w+1) (2 \beta_{BLL}-1)+2 (\beta_{BBL}-1) x_{3}+2 (\beta_{TBL}-1) x_{4}-2 \alpha\Big]\,,\label{phi'}\\
\frac{d\alpha}{dN}&=&2 \beta_{TLL}\phi^2-\alpha (2 \alpha+2 x_{3}+2 x_{4}+3 w+3)+2 \beta_{TTL}x_{1} \phi+\frac{2 \beta_{TBL} x_{3} x_{4} \phi}{x_{1}}\,,\label{alpha'}\\
\frac{d\theta}{dN}&=&-\theta \left(3 (w-1)-6 (w+1) \beta_{LL}+2 x_{3}+2 x_{4}+\frac{2 y_{1}}{y_{2}}\right)-\frac{x_{3} y_{1} \phi}{\alpha_{TB} y_{2}}-2 \alpha \left(\theta+\frac{y_{1}}{y_{2}}-3\right)\,.\label{theta'}
\end{eqnarray}
Additionally, one can use the Friedmann constraint (\ref{Fcons}) to reduce the above system to a 9-dimensional one. In the following sections we will explore the dynamical system of different kind of matter coupled theories of gravity which can be obtained from our approach.

\subsection{Specific model: $f(T,B,L_m)=\tilde{f}(-T+B,L_m)=\tilde{f}(R,L_m)$ gravity}
From our action it is possible to recover a very interesting model which comes from the curvature approach. As we discussed before, this is possible due to the fact that the function also depends on the boundary term $B$ so that it is always possible to reconstruct theories which contains the scalar curvature $R$. In this sense, one can construct a non-minimally coupled theory between the Lagrangian matter and the scalar curvature, explicitly by taking the function being
\begin{eqnarray}
f(T,B,L_m)=\tilde{f}(-T+B,L_m)=\tilde{f}(R,L_m)\,.\label{fRLm}
\end{eqnarray}
This kind of models was first proposed in \cite{Harko:2010mv} where the authors suggested that in general, this theory has some extra terms in the geodesic equation. The complete analysis of the dynamical system for flat FLRW for this model was studied in \cite{Azevedo:2016ehy}. From our full dynamical system, it is possible to recover the same dynamical system equations reported in the former paper. Let us recall here again that in general, the signature of the metric for curvature theories as $f(R)$ gravity is usually taken as the opposite as in teleparallel theories of gravity and hence the paper \cite{Azevedo:2016ehy} is written in another signature notation compared to our notation. The important of this notation issue is that the scalar curvature, scalar torsion and also the boundary term will have a minus sign of difference with respect to our case. Hence, in their notation Eqs.~(\ref{x1'})-(\ref{theta'}) will have a minus of difference in all those quantities. Therefore, to recover the same dynamical systems found in \cite{Azevedo:2016ehy} we need to change $R\rightarrow -R$ (and of course $T\rightarrow -T$ and $B\rightarrow -B$) which makes that the important derivatives appearing in the dimensionless variables become
\begin{eqnarray}
f_{B}=-f_{R}\,,\quad f_{T}=f_{R}\,,\quad f_{BB}=f_{TT}=-f_{TB}=f_{RR}\,.\label{fff}
\end{eqnarray}
In this case, some dimensionless variables can be reduced. It is possible to connect our dimensionless variables to the dimensionless variables used in the mentioned paper by working with the variables
\begin{eqnarray}
y=2 y_1-1\,,\quad  y_2=\frac{1}{2}\,,\quad x=2(x_1+x_2)=-2(x_3+x_4)\,,\quad \tilde{z}=2z\,,\quad \tilde{\phi}=2\phi=-2\alpha\,,\quad \tilde{\theta}=2\theta\,,\label{variabless}
\end{eqnarray}
where tildes represent the variables chosen in \cite{Azevedo:2016ehy}. By replacing (\ref{fff}) and (\ref{variables}) in our dynamical system we directly find that the corresponding dynamical system becomes a 5-dimensional one explicitly given by
\begin{eqnarray}\label{systemfR}
dx \over dN&= &x\left[x-y+\tilde{\phi}\left(1+\beta_{BBL}\right)\right] -1 -y -3\tilde{z} + \tilde{\theta} \nonumber\\ 
&&+ \tilde{\phi}\left[3(1+w)\left(2\beta_{BLL}+1\right)-y\right]\,, \\
dy \over dN &= & -\frac{x}{2 \alpha_{TB}}-y \left(\frac{x}{2 \alpha_{TB}}+2 y-4\right) \,,\\
d\tilde{z} \over dN &= & \tilde{z}(x+\tilde{\phi}+2(2-y))-\tilde{\theta}-\frac{x}{2\alpha_{TB}}(1+y)\,, \\
d\tilde{\phi} \over dN &= & \tilde{\phi}\left[x\left(1-\beta_{BBL}\right)- 3(1+w)\left(2\beta_{BLL}+1\right)+\tilde{\phi}\right] \,,\\
d\tilde{\theta} \over dN &= & \tilde{\theta} \left((6 (w+1) \beta_{LL}+x-2 y-3 w+1)+\tilde{\phi}\right)+\frac{x (y+1) \tilde{\phi}}{2 \alpha_{TB}}\,.
\end{eqnarray}
The above equations are the same reported in \cite{Azevedo:2016ehy} for $f(R,L_m)$ gravity if one changes the variables $\beta_{BBL},\beta_{BLL},\alpha_{TB}$ and $\beta_{LL}$ accordingly. In this theory, it is possible to reconstruct different interesting gravity-matter coupled models as for example standard non-minimally curvature-matter coupled models where $f(R,L_m)=f_1(R)+f_2(R)L_m$  which has been studied in the literature (see \cite{Bertolami:2007gv}). In \cite{An:2015mvw}, it was studied the dynamical system for some of those models. To find all the most important details regarding the above dynamical system for the former model and also for other more general models in $f(R,L_m)$ gravity, see \cite{Azevedo:2016ehy}.

\subsection{Specific model: $f(T,B,L_m)=\tilde{f}(T,L_m)$ gravity}
Let us now introduce a new theory of gravity based on an arbitrary function $f$ which depends on $T$ and $L_m$ only. As $f(T)$ was motivated by $f(R)$ gravity, $f(T,L_m)$ gravity is somehow, the teleparallel version of $f(R,L_m)$ discussed in the previous section. Different particular cases of this theory have been studied in the past. Let us first derive the full dynamical system for the $f(T,L_m)$ gravity and then study some particular theories. The Friedmann equation (\ref{Fcons}) for this model reads
\begin{eqnarray}
z=1\,.\label{z1}
\end{eqnarray}
In this case, $y_1=y_2=x_{1}=x_{2}=x_{3}=\phi\equiv 0$, so one needs to be very careful with the general dynamical system (\ref{x1'})-(\ref{theta'}) since some of these equation will be also identically zero. Let us clarify here the way that one needs to proceed to find the correct dynamical system. There are two ways to find out the correct dynamical system for an specific model. Let us discuss how to proceed with the model that we are interested here, i.e., where $f=f(T,L_m)$. The first way to proceed is using the full dynamical system described by (\ref{x1'})-(\ref{theta'}). If one directly replaces $f=f(T,L_m)$ in the full dynamical system, there will be some expressions that are indeterminate or directly zero, for example terms like $y_1/y_2$ or terms divided by $x_2$. Hence, one first needs to replace back all the original definitions of the dimensionless variables and after doing that, one can restrict $f=f(T,L_m)$. By doing that, several equations are directly satisfied. Indeed, one can verify that Eqs.~(\ref{x1'}), (\ref{x3'}), (\ref{y1'}), (\ref{y2'}) and  (\ref{phi'}) are identically zero, as expected. Then, for all the remaining equations, one needs to introduce again the dimensionless variables needed (in this case $x_4, \alpha$ and $\theta$). A second approach is to directly assume $f=f(T,L_m)$ in the Friedmann equations (\ref{eq1})-(\ref{eq2}) and then introduce the same dimensionless variables that we defined. By doing that, we arrive of course at the same dynamical system as the first approach. We will implement the first procedure in this work. Eq.~(\ref{x2'}) gives us a constraint for the variables, namely
\begin{eqnarray}
\frac{y_1}{y_2}= -2 \alpha-\theta-2 x_4+3 z\,.\label{y1y2}
\end{eqnarray}
Let us here clarify again that even though $y_1=y_2\equiv 0$, the quotient $y=y_1/y_2=B/T=3+\dot{H}/H^2$ is clearly non-zero. If we replace the above equation and also use the Friedmann constraint (\ref{z1}), the remaining three Eqs. (\ref{x4'}), (\ref{alpha'}) and (\ref{theta'}) gives us the following set of equations,
\begin{eqnarray}
\frac{dx_4}{dN}&=&-\frac{1}{2 \alpha+\theta} \Big[2 \alpha \tilde{\beta}_{TTL} (2 \alpha+\theta)^2+x_{4} \left(4 \alpha^2 (6 \tilde{\beta}_{TTL}-\tilde{\beta}_{TLL}+3)+6 \alpha (2 \tilde{\beta}_{TTL} \theta+\theta+w+1)-3 (w+1) (2 \gamma_{LL}-1) \theta\right)\nonumber\\
&&+x_{4}^2 (-4 \alpha (\alpha_{TTT}-4 \tilde{\beta}_{TTL}-6)-2 (\alpha_{TTT}-4) \theta)-4 (\alpha_{TTT}-3) x_{4}^3\Big]\,, \label{x4'new}\\
\frac{d\alpha}{dN}&=& \alpha \Big[2 \alpha (\tilde{\beta}_{TLL}-2 \tilde{\beta}_{TTL}-1)-2 \tilde{\beta}_{TTL} \theta-2 (2 \tilde{\beta}_{TTL}+1) x_{4}-3 w-3\Big]\,,\label{alpha'new}\\
\frac{d\theta}{dN}&=&4 \alpha \theta+4 \alpha^2+\theta (2 \theta+6 (w+1) \gamma_{LL}-3 w-3)+2 x_{4} (2 \alpha+\theta)\,,\label{theta'new}
\end{eqnarray}
where for convenience we have introduced the following quantities
\begin{eqnarray}
\tilde{\beta}_{TLL}=\frac{f_{BL}^2}{f_{TL}^2}\beta_{TLL}= \frac{f_{T}f_{TLL}}{f_{TL}^2}\,,\quad \tilde{\beta}_{TTL}= \frac{T f_{TTL}}{f_{TL}}\,.
\end{eqnarray}
In the following section we will study some interesting cases that can be constructed from this theory.

\subsubsection{Nonminimal torsion-matter coupling $f(T,L_m)=f_1(T)+f_2(T)L_m$  }\label{sectionnn}
In this section we assume that the function takes the following form,
\begin{eqnarray}
f(T,L_m)=f_1(T)+f_2(T)L_m\,,
\end{eqnarray} 
where $f_1(T)$ and $f_2(T)$ are arbitrary functions of the torsion scalar. This model is an extension of $f(T)$ gravity, where an additional nonminimal coupling between the torsion and the matter Lagrangian is considered \cite{Harko:2014sja}. In this model we have that $\gamma_{LL}=\tilde{\beta}_{TLL}\equiv 0$. In~\cite{Carloni:2015lsa}, the dynamical system of this model was carefully studied. The authors used other dimensionless variables but one can verify that our dynamical system (\ref{x4'new})-(\ref{theta'new}) give rise to the the same dynamics. In that paper, the authors used a different energy density which is related to our as $\rho_{\rm new}=-2\rho$. The dimensionless variables used in \cite{Carloni:2015lsa} are given by 
\begin{eqnarray}
Y=\frac{f_{2}(T)}{12H^2 f_{2}'(T)}\,,\quad X=\frac{f_{1}(T)}{12H^2 f_{1}'(T)}\,, \quad \Omega=-\frac{\rho}{3 H^2}\,.
\end{eqnarray}
It is possible to show that our variables are related to them as follows
\begin{eqnarray}
Y=\frac{\theta }{2 \alpha }\,, \quad X=\frac{\theta -3 (w+1)}{2 \alpha +3 (w+1)}\,,\label{relation}
\end{eqnarray}
which gives us the following set of equations,
\begin{eqnarray}
\frac{dX}{dN}&=&-\frac{3 (w+1)  Q (X+1) (Y+1) (2 (W+1) X+1)}{P (X+1)-Q (2 W (Y+1)-X+Y)}\,, \label{11}\\
\frac{dY}{dN}&=&-\frac{3 (w+1)  (X+1) (Y+1) (P Y+2 Q Y+Q)}{P (X+1)-Q (2 W (Y+1)-X+Y)}\,,\\
X&=&-1-2\Omega Q (Y+1)\,,\label{33}
\end{eqnarray}
where 
\begin{eqnarray}
P=-\frac{T^2f_2''(T)}{f_1'(T)}\,,\quad Q=-\frac{Tf_2'(T)}{2f_1'(T)}\,, \quad W=\frac{Tf_1''(T)}{f_1'(T)}\,.
\end{eqnarray}
It can be shown that the Eqs.~(\ref{11})-(\ref{33}) are equivalent to our equations (\ref{x4'new})-(\ref{theta'new}) if the corresponding (\ref{relation}) is used properly. This for sure is a good consistency check that our equations are correct. The full study of the dynamical system (\ref{11})-(\ref{33}) was carried out in \cite{Carloni:2015lsa} where 6 different  kind of functions $f_1(T)$ and $f_2(T)$ were assumed. For some of those models, they found some critical points representing accelerating or decelerating solutions and also scaling solutions. For more details about all of this models and their dynamical analysis, see \cite{Carloni:2015lsa}.

\subsubsection{Exponential couplings for $f(T,L_m)$ gravity}
Now, let us study a new model where the function takes the following form
\begin{eqnarray}
f(T,L_m)=-\Lambda\exp\Big[-\frac{1}{\Lambda}\Big(T+L_m\Big)\Big]\,,\label{fexponentialT}
\end{eqnarray}
where $\Lambda$ is a positive cosmological constant. Let us take a look at this model further. In the limit where the argument is much less than one ($\frac{1}{\Lambda}(T+L_m)\ll 1$), if one expands up to first order in the argument, the function becomes
\begin{eqnarray}
f(T,L_m)\approx -\Lambda + T+L_m+\cdots\,,
\end{eqnarray}
hence in that limit, one recovers the TEGR plus matter case with a cosmological constant. Therefore, the function (\ref{fexponentialT}) is an interesting model to take into account. An analogous model was proposed in \cite{Harko:2010mv}, where instead of having $T$, the authors considered the scalar curvature $R$. The dynamical system of the former model was investigated in full detail in \cite{Azevedo:2016ehy}.\\
Under this theory, we directly find that $\tilde{\beta}_{TLL}=\alpha_{TTT}=1$ and by manipulating the definitions of the other quantities, $\beta_{LL}$ and $\tilde{\beta}_{TTL}$ can be written as
\begin{eqnarray}
\beta_{LL}=\frac{\alpha}{3(1+w)}\,,\quad \tilde{\beta}_{TTL}=-\frac{\alpha}{\theta}\,.
\end{eqnarray}
From the Friedmann constraint (\ref{z1}), $L_m=-2\rho$ and by using Eqs.~(\ref{rho}) and (\ref{TBR}) we directly find that for this model, the universe is always expanding as a De-Sitter one with a scalar factor being equal to
\begin{eqnarray}
a(t)\propto e^{\pm\frac{t}{2}\sqrt{\frac{\Lambda}{3}}}\,.
\end{eqnarray}
It is interesting to see that actually this model is very different to its analogous in $f(R,L_m)$. In the model  $f(R,L_m)=-\Lambda\exp[-(R+L_m)/\Lambda]$, it is not possible to directly find an unique scale factor which rules out the whole dynamic for the model. Hence, in that case, the dynamical system technique is very useful to check how the dynamics evolves on time. In our case, since the dynamics is always the same (described by the above equation) it is not important to study its dynamical properties since the solution is the standard De-Sitter universe. Therefore, the model described by an exponential coupling between $L_m$ and $T$ as it is given by (\ref{fexponentialT}) mimics a De-Sitter universe.

\subsubsection{Power-law couplings $f(T,L_m)$ gravity}
Let us consider another interesting new model that one can consider from our approach where the function takes the following form
\begin{eqnarray}
f(T,L_m)=M^{-\epsilon}( T+L_m)^{1+\epsilon}\,,\label{powerlawfT}
\end{eqnarray}
where $\epsilon$ is a constant and $M$ is another constant which represents a mass characteristic scale. In this case, up to first order in $\epsilon$, the expansion of the above function becomes
\begin{eqnarray}
f(T,L_m)\approx T+L_m+\epsilon  (T+L_m) \log\Big[\frac{T+ L_m}{M}\Big]\,,
\end{eqnarray}
so that since $\epsilon$ is assumed to be very small (comparable with $T$ and $L_m$), the above model could represents a small deviation of the standard TEGR plus matter case. For this model we find that
\begin{eqnarray}
\tilde{\beta}_{TLL}=\alpha_{TTT}=1-\epsilon^{-1}\,,\quad \tilde{\beta}_{TTL}=-\frac{(\epsilon -1) \alpha }{\epsilon  \theta }\,,\quad \beta_{LL}=\frac{\alpha}{3(1+w)}\,.
\end{eqnarray}
Similarly as we did in the previous section, from the Friedmann constraint (\ref{z1}), by replacing $L_m=-2\rho$ and by using Eqs~(\ref{rho}) and (\ref{TBR}) we find the following equation for the scale factor,
\begin{eqnarray}
\left(\rho_0+3a^{3 w+1} \dot{a}^2\right)^{\epsilon } \left(\rho_0-3 (2 \epsilon +1) a^{3 w+1} \dot{a}^2\right)&=&0\,,\label{eqqm}
\end{eqnarray}
 which gives us two different types of scale factors. One can directly check that if $\epsilon=0$, the above equation is reduced to the standard TEGR plus matter case, namely $3H^2=\rho$. For the specific case where $\epsilon=-1/2$, we must need $\rho=0$, so that this special case is not a reliable model. There is no point on going further with the dynamical system of this model since the equation can be directly solved for the scale factor. The above equation depends on the power-law parameter $\epsilon$. For negatives values of $\epsilon$, the only possibility is that the second bracket is zero whereas for positives values of $\epsilon$, there will be two kind of possible scale factor. This is again different as the case $f(R)=M^{-\epsilon}(R+L_m)^{-\epsilon+1}$ studied in \cite{Azevedo:2016ehy}. Our model seems to be simpler than the former one due to the fact that $T$ only contains derivatives of $a(t)$ and not second derivatives as $R$.\\ 
 Let us know explore what kind of solutions we have from our power-law model. The first type can be obtained by assuming that the first bracket is zero, which is only valid for $\epsilon>0$ giving us the following scale factor,
 \begin{eqnarray}
a_{\pm}(t)&=& \left(\frac{3}{4}\right)^{\frac{1}{3 w+3}} \left(\pm i\sqrt{\rho_0} t (w+1)\right)^{\frac{2}{3 (w+1)}}\,, \ \ \textrm{where} \ \ \epsilon>0\,,\label{aa}
 \end{eqnarray}
where for simplicity we have chosen that the integration constant is zero. Let us clarify here that this scale factor will rule out the dynamic only for $\epsilon>0$. The scale factor must be real and positive so we must ensure that the imaginary term disappears. This is possible for some values of $w$. If one assumes that $w>-1$, for the solution $a_{+}(t)$, the state parameter must satisfy $w_{+}=\frac{1}{6 k}-1$ for any positive integer number $k$ whereas for the solution $a_{-}$, the state parameter must be $w_{-}=\frac{11}{6 k}-1$ to ensure a positive real value of $a_{\pm}(t)$. Moreover, for these two solutions, $\dot{a}_{\pm}>0$ and $\ddot{a}_{\pm}>0$ for both $w_{\pm}$, so this solution could describe an accelerating expanding universe for those specific values of $w_{\pm}$. However, only the solution $a_{-}$ with $w_{-}=11/6\approx 1.8$ represents power-law expanding accelerating universes without evoking exotic matter. \\
Additionally, Eq.~(\ref{eqqm}) can be solved by letting the second bracket equal to zero, which is valid for all $\epsilon\neq -1/2$, yielding
 \begin{eqnarray}
 a(t)= \left(\frac{3 \rho_0}{4 (2 \epsilon +1)}\right)^{\frac{1}{3 (w+1)}}t^{\frac{2}{3 (w+1)}}\,,\label{a}
 \end{eqnarray}
where again for simplicity we have assumed that the integration constant is zero. This solution is very similar to (\ref{aa}) but now the parameter $\epsilon$ plays a role in the dynamics of the universe. Let us again consider the case where $w>-1$ for studying this solution. For $\epsilon>-1/2$, the scale factor and its derivatives are always positive so that the scale factor will mimic a power-law accelerating universe. For $\epsilon<-1/2$, we need to impose that
\begin{eqnarray}
w=-1+\frac{1}{6k}\,, \ \ \textrm{where} \ \ \ k\in \mathbb{Z}^{+}\,,
\end{eqnarray}
otherwise the scale factor would be negative. Moreover, all the derivatives of the scale factor would be also positive if $w$ satisfy the above condition .Hence, only special cases of $w$ will give rise to viable models when $\epsilon<-1/2$. Further, all those models are in the regime $-1<w<0$ which represents exotic kind of matter. Thus, cases with $\epsilon<-1/2$ needs exotic matter to represent accelerating expanding universes. Additionally, we can conclude that for $\epsilon>1/2$, the power-law $f(T,L_m)$ will mimic power-law accelerating universes without evoking exotic matter.

\subsection{Specific model: $C_1 T+f(B,L_m)$ gravity }
In this section we will study the case where the function takes the following form
\begin{eqnarray}
f(T,B,L_m)=C_1 T+\tilde{f}(B,L_m)\,,
\end{eqnarray}
where $C_1$ is a constant and the function $\tilde{f}(B,L_m)$ depends on both the boundary term and the matter Lagrangian. The first term represents the possibility of having TEGR (or GR) in the background when we set $C_1=1$. If this term does not appear in the function, it is not possible to recover GR since one cannot construct GR from $\tilde{f}(B,L_m)$ gravity. This kind of theories have not been considered in the past, but there are some studies for the specific case $\tilde{f}(B,L_m)=f(B)+L_m$, which is known as $f(B)$ gravity \cite{Paliathanasis:2017flf,Bahamonde:2016grb,Bahamonde:2016cul}. The full dynamical system (\ref{x1'})-(\ref{theta'}) is simplified since $x_1=x_3=x_4=\alpha\equiv 0$ which implies that Eqs.~(\ref{x1'}), (\ref{x3'}), (\ref{x4'}) and (\ref{alpha'}) are also automatically zero. Hence, in our variables, this theory is a 5-dimensional dynamical system given by
\begin{eqnarray}
\frac{dx_2}{dN}&=& -6 (w+1)\beta_{BLL} \phi +\theta +3 (w+3) \phi+x_2 (6-2 \tilde{\beta}_{BBL} \phi) -3-\frac{y_1 (x_2+\phi -1)}{y_2}\,,\label{x2'f(LB)}\\
\frac{dy_1}{dN}&=&-\frac{ x_2 (y_1+\beta_{BB} y_2)+y_1 (2 y_1-6 y_2+\phi)}{y_2}\,,\label{y1'f(LB)}\\
\frac{dy_2}{dN}&=&-(x_2+\phi)\,,\label{y2'f(LB)}\\
\frac{d\phi}{dN}&=&2(3 \beta_{BLL}-1)\phi(w+1)+2 \tilde{\beta}_{BBL} x_2 \phi\,,\label{phi'f(LB)}\\
\frac{d\theta}{dN}&=&3 \theta  \Big(2 (w+1) \beta_{LL}-w+1\Big)+\frac{ \beta_{BB} x_2 \phi-2 \theta  y_1}{y_2}\,,\label{theta'f(LB)}
\end{eqnarray}
where for simplicity we have introduced the following quantities
\begin{eqnarray}
\tilde{\beta}_{BBL}=\beta_{BBL}\frac{f_{TB}}{f_{BB}}=\frac{f_{T} f_{BBL}}{f_{BL}f_{BB}}\,,\quad  \beta_{BB}=\frac{f_{B}}{T f_{BB}}\,.
\end{eqnarray}
Let us now concentrate on a specific model based on the boundary term non-minimally coupled with the the matter Lagrangian where the function takes the following form
\begin{eqnarray}
f(T,B,L_m)=C_1 T+f_1(B)+f_2(B)L_m\,,
\end{eqnarray}
where $C_1$ is a constant and $f_1(B)$ and $f_2(B)$ are functions which depends on the boundary term $B$. This case is analogous to the one studied in Sec.~\ref{sectionnn}, but the dynamical system is more complicated to deal since it is a 5 dimensional one. The aim of this section is to study some specific cases that can be constructed from the above model. \\
Let us further study the case where the functions are a power-law type given by
\begin{eqnarray}
f_1(B)=C_5 B^{s}\,, \quad f_2(B)=(C_4+C_3 B^{q})L_m\,,\label{f1f2}
\end{eqnarray}
where $C_3,C_4,C_5,q$ and $s$ are constants. Since we are interested on studying non-trivial couplings between $B$ and $L_m$ we will assume that $C_3\neq0$. We directly find that $\beta_{LL}=\beta_{BBL}=0$. It can be proved that for this model, the dynamical system can be reduced from 5 dimensional to a 4 dimensional one. For this case, the dynamical system is difficult to study. However, if one assumes that the exponents are related as
\begin{eqnarray}
q=1-s\,,\label{qs}
\end{eqnarray}
the system becomes easier to work since it becomes a 3 dimensional dynamical system. Then, we will split the study depending on different cases which depends on the constants. 
\subsubsection{$f(T,B,L_m)=C_1 T+C_5 B^{s}+(C_4+C_3 B)L_m$}
Let us first study a very special case where $q=1$ in \eqref{f1f2} giving us a linear coupling between the boundary term  $B$ and the matter Lagrangian $L_m$. This model will depend on the power-law parameter $s$ and also on the constants $C_3,C_4$ and $C_5$. In this model, one can relate two dynamical dimensionless variables with the other ones making it a 3 dimensional one. In our case, we will replace $\theta$ and $y_1$ as follows
\begin{eqnarray}
\theta&=&-\frac{3 C_{1} (w+1) \phi (2 \phi+2 x_{2}-1) \left[C_{3}  \left(\frac{2C_{1} (\phi-3 (w+1) y_{2})}{3C_{5} s (w+1)}\right)^{\frac{1}{s-1}}+C_{4}\right]}{C_{3} C_{5}   (s-1) (w+1) \left(\frac{2\cdot 3^{1/s}C_{1} (\phi-3 (w+1) y_{2})}{C_{5} s (w+1)}\right)^{\frac{s}{s-1}}+2 C_{1} C_{4} \phi}\,,\\
y_1&=&\frac{C_{3} (2 \phi+2 x_{2}-1) \left[C_{1} \phi \left(\frac{C_{1} 2^s (\phi-3 (w+1) y_{2})}{C_{5} s (w+1)}\right)^{\frac{1}{s-1}}-C_{5} s (w+1) \left(\frac{2C_{1} (\phi-3 (w+1) y_{2})}{C_{5} s (w+1)}\right)^{\frac{s}{s-1}}\right]}{2C_{3} C_{5}  (s-1) (w+1) \left(\frac{2C_{1} (\phi-3 (w+1) y_{2})}{C_{5} s (w+1)}\right)^{\frac{s}{s-1}}+4 C_{1} C_{4} 3^{\frac{1}{s-1}} \phi}\,.
\end{eqnarray} 
Thus, it is possible to replace the above equations into Eqs.~(\ref{y2'f(LB)})-(\ref{phi'f(LB)}) in order to reduce the dimensionality of the dynamical system for this model. By doing that, we find that the model only has one critical point given by
\begin{eqnarray}
P:\quad \big(x_2,y_2,\phi\big)=\Big(0,\frac{s}{6 (s-1)},0\Big)\,,
\end{eqnarray}
which depends on the power-law parameter $s$. The case $s=1$ can be discarded since a linear combination of the boundary term does not affect the field equations. It is easily to see that the effective state parameter for this critical point is always $-1$, hence this critical point always represents acceleration. To find out about the stability of this point, one needs to check the eigenvalues evaluated at $P$. There are three different eigenvalues given by
\begin{eqnarray}
\Big\{-3(1+w)\,, -\frac{3}{2}\pm \frac{3 \sqrt{(8-7 s) s}}{2 s}\Big\}\,.
\end{eqnarray}
One can directly see that when  $1<s\leq 8/7$ and $w>-1$, the critical point $P$ is stable.
\subsubsection{$f(T,B,L_m)=C_1 T+(C_4+C_3 B^{q})L_m$}
\begin{figure}[H]
	\centering
	\includegraphics[width=0.5\textwidth]{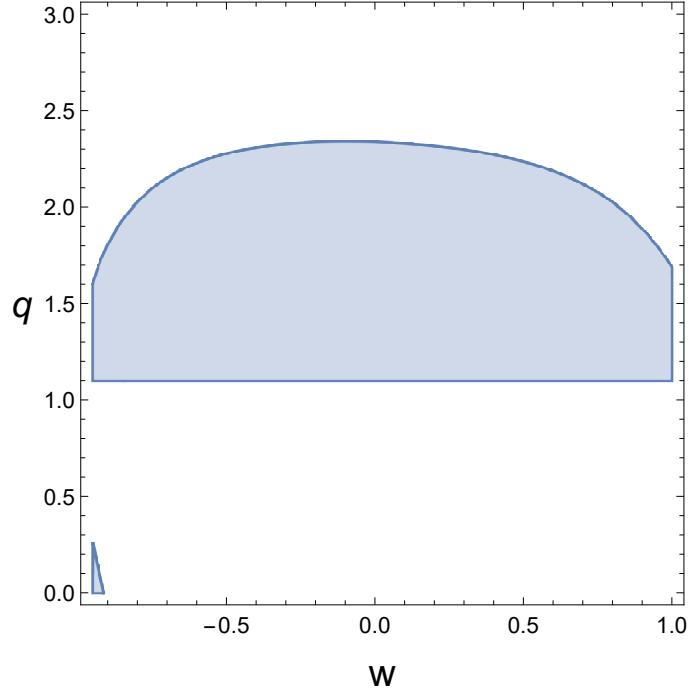}
	\caption{Region plot for the state parameter $w$ and the power-law parameter $q$ for the model described by (\ref{f1f2}) with $C_5=0$. The figure represents the regions where the point (\ref{P}) is stable. The blank regions represent the regions where the point is a saddle one.}
	\label{Fig5}
\end{figure}

Let us now consider the case where $C_5=0$ in \eqref{f1f2}. This model represents the case where $f_1(B)=0$. Let us also assume that $q\neq 1$ to do not have the same model as the previous case. For this case, it is possible to relate the terms $\tilde{\beta}_{BBL}$ and $\beta_{BB}$ with the dynamical variables as follows
\begin{eqnarray}
\tilde{\beta}_{BBL}=-\frac{3 (w+1)}{2 \phi }\,,\quad \beta_{BB}=-\frac{2 \theta+3 (w+1) (2 \phi-1)+6 (w+1) x_{2}}{2 (q-1) \phi}\,.\label{case1BB}
\end{eqnarray}
Moreover, the dynamical system can be reduced from 5D to 3D since some of the variables are directly related, namely
\begin{eqnarray}
y_1=-\phi -\frac{\theta}{3 (w+1)}-x_2+\frac{1}{2}\,,\quad y_2=\frac{\phi}{3 (w+1)}\,.\label{case1B2}
\end{eqnarray}
By replacing (\ref{case1BB}) and (\ref{case1B2}) in the dynamical system (\ref{x2'f(LB)})-(\ref{theta'f(LB)}) we find that this system is reduced as follows
\begin{eqnarray}
\frac{dx_2}{dN}&=&\frac{\theta (4\phi-2)+3 \left(4 (w+2)\phi^2-(3 w+5)\phi+w+1\right)+x_{2} (2 \theta+6 (3 w+5)\phi-9 (w+1))+6 (w+1) x_{2}^2}{2\phi}\,,\\
\frac{d\phi}{dN}&=&-3 (w+1) (\phi+x_{2})\,,\\
\frac{d\theta}{dN}&=&\frac{3 (w+1)}{\phi} \left(\frac{3 (w+1) x_{2}}{q-1}-2 \theta\right) \left(-\phi-\frac{\theta}{3 (w+1)}-x_{2}+\frac{1}{2}\right)-3 (w-1) \theta\,.
\end{eqnarray}
This dynamical system has only one critical point given by
\begin{eqnarray}
(x_2{}_{*},\phi_{*},\theta_{*})=\Big(-\frac{q (w+1)}{2 q+w-1}, \frac{q (w+1)}{2 q+w-1},\frac{3 w+3}{2-2 q}\Big)\,,\label{P}
\end{eqnarray}
where we have assumed that $q\neq (1-w)/2$. This point of course depends on the parameters $q$ and $w$. For this point, there is acceleration when 
\begin{eqnarray}
\frac{q+w}{1-q}<-\frac{1}{3} \Longrightarrow q>\frac{1}{2} (-3 w-1)\,,
\end{eqnarray}
where we have assumed that $w>-1$. Further, for the dust case $w=0$ we can see that this point requires $q>-1/2$ to represent and accelerating universe. It is also possible to check that there are three Eigenvalues associated with this point. Those Eigenvalues are very long to present here but Fig.~\ref{Fig5} represents a region plot where the point is stable. Note that besides of the values of $q$ and $w$, this point is never unstable.

\subsubsection{$f(T,B,L_m)=C_1 T+C_5 B^{s}+(C_4+C_3 B^{1-s})L_m$}
\begin{figure}[H]
	\centering
		\includegraphics[width=0.5\textwidth]{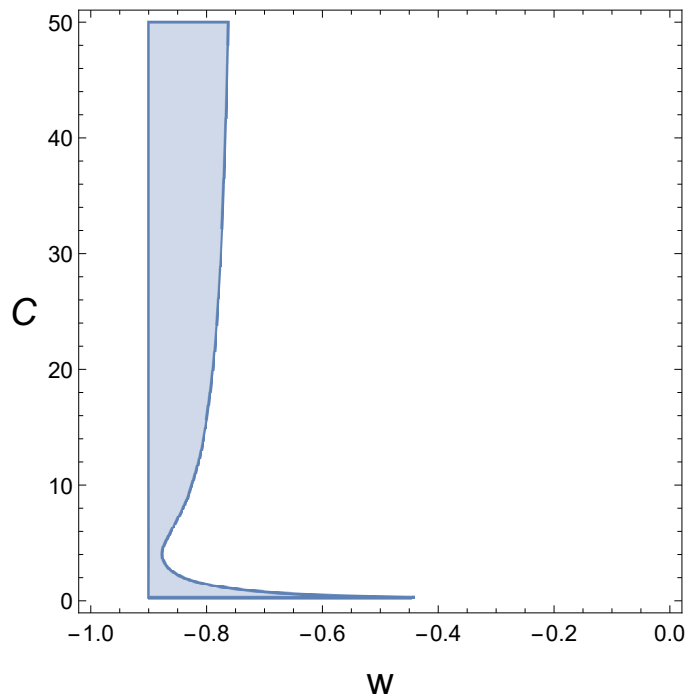}
	\caption{Region plot for the critical point $P_2$ for the model $s=-2$ for the constants $C_1=C_4=1$ and $C=C_5C_3$. The figure is representing the regions where the critical point for that model (see Table~\ref{Table1}) is  stable. The point is never unstable. All the blank regions represents the regions where the point is a saddle point.}
	\label{Fig9}
\end{figure}

Let us now assume the case where $q=1-s$ and the constant $C_5\neq0$ which is a more generic model which has an additional boundary power-law contribution. As we have studied in the previous section, we can again reduce the dynamical system as a 3-dimensional one. However, the models is much more complicated than the previous two models. The dynamics of the model highly depends on the parameter $s$. We can relate the terms $\tilde{\beta}_{BBL}$ and $\beta_{BB}$ with the dimensionless variables but now those quantities are very long for a generic $s$. Moreover, those terms make the dynamical system very long and difficult to treat for any $s$. One can also relate two dimensionless variables with the other ones. In this case, we will choose to work with the variables $(y_2,\theta,\phi)$ since the dynamical system is slightly easier to work with them. The variables $x_2$ and $y_1$ are then given by
\begin{eqnarray}
x_2&=&\frac{1}{6 (w+1) \phi (2 C_{1} C_{4} \phi-6 C_{1} C_{4} (w+1) y_{2}+3 C_{3} C_{5} s (w+1))}\Big[\phi (-2 \theta(2 C_{1} C_{4} \phi+3 C_{3} C_{5} (2 s-1) (w+1))\nonumber\\
&&-3 (w+1) (2 \phi-1) (2 C_{1} C_{4} \phi+3 C_{3} C_{5} s (w+1)))-6 (w+1) y_{2} \Big(\theta\left(3 C_{3} C_{5} (s-1)^2 (w+1)-2 C_{1} C_{4} \phi\right)\nonumber\\
&&-3 C_{1} C_{4} (w+1) \phi (2 \phi-1)\Big)\Big]\,,\\
y_1&=&-\frac{s y_{2} (2 \theta+3 (w+1) (2 \phi-1)+6 (w+1) x_{2})}{2 (\phi+3 (s-1) (w+1) y_{2})}\,.
\end{eqnarray}
It is possible to write down the dynamical system for any generic $s$ but it is very long a cumbersome to present it here. Moreover, the critical points highly depend on the parameter $s$ and it is not possible to obtain all the possible critical points for any arbitrary $s$. Hence, we will only study some particular models. We will concentrate only on models with integer values of $s$. Table~\ref{Table1} represents various models with their critical points, effective state parameter and their acceleration regime. In general, for all the critical points for those models, it is possible to have acceleration for the dust case $w=0$.
%\clearpage
%\begin{center}
 \begin{table}[t]
 	\captionsetup{justification=raggedright, labelsep=space}
	\scalebox{0.65}{\begin{tabular}{|c|c|c|c|c|}
			\hline
			$s$ & Model & $(y_2,\theta,\phi)$ & $w_{\rm eff}$ & Acceleration\\ 
			\hline \hline
			$\displaystyle s=2$ & $ \displaystyle C_1 T+C_5B^2 + \big( \frac{C_3}{B}+C_4\big)L_m$   &   $\big(\displaystyle\frac{1}{9-3 w},\displaystyle\frac{3 (w+1)}{4},\displaystyle\frac{w+1}{3-w}\big)$ & $\displaystyle\frac{w-1}{2}$ & $w<1/3$ \\ 
			\hline
			$\displaystyle s=3$ & $\displaystyle C_1 T+C_5B^3 + \big(\frac{C_3}{B^2}+C_4\big)L_m$   &   $\big(-\displaystyle\frac{2}{3 (w-5)},\displaystyle\frac{w+1}{2},-\displaystyle\frac{2 (w+1)}{w-5}\big)$ & $\displaystyle\frac{w-2}{3}$ & $w<1$ \\ 
			\hline
			$\displaystyle s=4$ & $\displaystyle C_1 T+C_5B^4 + \big(\frac{C_3}{B^3}+C_4\big)L_m$   &   $\big(\displaystyle\frac{1}{7-w},\displaystyle\frac{3 (w+1)}{8},-\displaystyle\frac{3 (w+1)}{w-7}\big)$ & $\displaystyle\frac{w-3}{4}$ & $w<5/3$ \\ 
			\hline
			$\displaystyle s=5$ & $\displaystyle C_1 T+C_5B^5 + \big(\frac{C_3}{B^4}+C_4\big)L_m$   &   $\big(-\displaystyle\frac{4}{3 (w-9)},\displaystyle\frac{3 (w+1)}{10},-\displaystyle\frac{4 (w+1)}{w-9}\big)$ & $\displaystyle\frac{1}{20} (3 w-7)$ & $w<1/9$ \\ 
			\hline
			$\displaystyle s=-1$ & $\displaystyle C_1 T+\frac{C_5}{B} + \big(C_3B^2+C_4\big)L_m$   &   $\big(\displaystyle\frac{2}{3 (w+3)},-\displaystyle\frac{3}{2} (w+1),\displaystyle\frac{2 (w+1)}{w+3}\big)$ & $-(w+2)$ & $w>-5/3$ \\ 
			\hline
			\multirow{11}{*}{$\displaystyle s=-2$} &\multirow{11}{*}{$ \displaystyle T+\frac{C_5}{B^2} + \big(C_3B^3+1\big)L_m$}   &  \multirow{1}{*}{$P_{1}$}  & \multirow{7}{*}{$\displaystyle\frac{98 +1095 C_3C_5 (w+1)-\Delta}{300 C_3C_5}$} &$C_5C_3>0$ $\&$ $w<\displaystyle\frac{-3980 C_5C_3-539}{4380 C_5C_3+735}$\\ 
			& &\multirow{2}{*}{$\big(-\displaystyle\frac{98 +15  C_{3} C_{5} (23 w+43)+\Delta}{336  (5 w+1)},$} &  & $\lor\ -\displaystyle\frac{49}{438}\leq C_5C_3<0$ $\&$ $w<-\displaystyle\frac{3980 C_5C_3+539}{4380 C_5C_3+735}$ \\
			& &\multirow{2}{*}{$-\displaystyle\frac{3(1+w)(154 +45  C_{3}  C_{5} (73 w+53)-3\Delta)}{2240},$}
			& & $\lor\ -\displaystyle\frac{49}{292}\leq C_5C_3<-\displaystyle\frac{49}{438}$ $\&$ $w<-\displaystyle\frac{58035 C_5C_3+560  \sqrt{-3(438 C_5C_3-49)}+2254}{79935 C_5C_3}$ \\
			& & 	
			\multirow{2}{*}{$-\displaystyle\frac{(1+w)(98 +15  C_{3} C_{5} (11-137 w)+\Delta)}{112 (5 w+1)}\big)$} & & $\lor\ \Big(-\displaystyle\frac{49 \left(2 \sqrt{474}+61\right)}{7300}<C_{5}C_3<-\displaystyle\frac{49}{292}\&$ \\
			&&&& $-\displaystyle\frac{3980 C_{5}C_3+539}{4380 C_{5}C_3+735}<w\leq -\displaystyle\frac{58035 C_{5}C_3+560 \sqrt{3} \sqrt{-438 C_{5}C_3-49}+2254}{79935 C_{5}C_3}\Big)$\\
			& & & & $\lor\ \Big(\displaystyle\frac{49 \left(2 \sqrt{474}-61\right)}{7300}<C_{5}C_3<-\displaystyle\frac{49}{438}\& $\\
			&&&& $-\displaystyle\frac{58035 C_{5}C_3-560 \sqrt{3} \sqrt{-438 C_{5}C_3-49}+2254}{79935 C_{5}C_3}\leq w<-\displaystyle\frac{3980 C_{5}C_3+539}{4380 C_{5}C_3+735}\Big)$\\ \cline{3-5}
			%new part
			&&  \multirow{1}{*}{$P_{2}$}  & \multirow{4}{*}{$\displaystyle\frac{98 +1095 C_3C_5 (w+1)+\Delta}{300 C_3C_5}$} & $-\displaystyle\frac{49}{438}\leq C_5C_3<0$ \\ 
			& &\multirow{1}{*}{$\big(\displaystyle\frac{-15C_5C_3 23 w+43)-98+\Delta}{336  (5 w+1)},$} &  & $\lor\ -\displaystyle\frac{49}{292}<C_5C_3\leq -\displaystyle\frac{98 \sqrt{474}+2989}{7300}\ \& \ w>\displaystyle\frac{-3980 C_5C_3-539}{4380 C_5C_3+735}$\\
			& &\multirow{1}{*}{$-\displaystyle\frac{3(1+w)(154 +45  C_{3}  C_{5} (73 w+53)+3\Delta)}{2240},$}
			& &   \\
			& & 	
			\multirow{1}{*}{$\displaystyle\frac{15 C_5C_3 (137 w-11)-98+\Delta)}{112 (5 w+1)}\big)$} & & $\lor \ C_5C_3\leq \displaystyle\frac{-98 \sqrt{474}-2989}{7300} \ \& \ w<\displaystyle\frac{-3980C_5C_3-539}{4380 C_5C_3+735}$  \\ \hline
	\end{tabular}} 
\caption{Acceleration and effective state parameter of the critical points for different models which depend on $s$. The model studied is described by $f(T,B,L_m)=C_1T+C_5 B^{s}+(C_4+C_3 B^{1-s})L_m$ and  $\Delta=\sqrt{225 (C_5C_3)^{2} (73 w+53)^2+2940 C_5C_3 (23 w+43)+9604}$.}
	\label{Table1}
\end{table}
%\end{center}
 It is also important to mention that for $s\geq 2$, all the models have only one critical point with the possibility of describing acceleration depending on the state parameter $w$. It can be proved that the critical points in the models $s=3,4,5$ (there is only one critical point for each model) are always saddle points. The model $s=2$ can be either a saddle point or an unstable point. Hence, for all of positive models of $s$, the critical points cannot be stable.
When negatives values of $s$ are considered, the system becomes more complicated. For $s\leq-3$, the dynamical system becomes highly complicated to analyse.  
 For the case $s=-1$,  the critical point is either a saddle or an unstable point so it cannot be stable. Moreover, for the dust case ($w=0$), the critical point for the model $s=-1$ is always unstable spiral. For the case $s=-2$, there are two critical points $P_1$ and $P_2$ (see Table~\ref{Table1}). The critical point $P_2$ is either a saddle point or stable whereas the point $P_1$ is always a saddle point. Fig.~\ref{Fig9} represents the regions where the point $P_2$ is stable. It is important to mention that only the term $C=C_3C_5$ appears in the Eigenvalues so that it is possible to make 2D region plots for the model. In this figure, it was considered the case $C_1=C_4=1$ which is equal to consider the standard General Relativity plus matter model in the background.
\section{Conclusions}\label{sec:4}
\begin{figure}[h]
	\centering
	\begin{tikzpicture}
	\matrix (m) [matrix of math nodes,row sep=8em,column sep=13em,minimum width=3em]
	{ \mbox{}   & f(T,B,L_m)  & \mbox{}\\
		\boxed{f(R,L_m)} & C_1T+f(B,L_m)   & \boxed{f(T,L_m)}\\
		f_1(R)+f_2(R)L_m & C_1T+f_1(B)+f_2(B)L_m   & f_1(T)+f_2(T)L_m\\
		\boxed{f_1(R)+L_m} & C_1T+f_1(B)+L_m   & \boxed{f_1(T)+L_m}\\
		&  \boxed{\text{GR \& TEGR}} & \mbox{} \\};
	\path[-stealth,every node/.style={sloped,anchor=north}]
	(m-1-2) edge node [above] {$f=f(T,L_m)$} (m-2-3)
	(m-1-2) edge node [above] {$f=f(-T+B,L_m)$} (m-2-1)
		(m-1-2) edge node [right,rotate=90] {$f=C_1T+f(B,L_m)$} (m-2-2)
			(m-2-2) edge node [right,rotate=90] {$f=f_1(B)+f_2(B)L_m$} (m-3-2)
				(m-2-1) edge node [right,rotate=90] {$f=f_1(R)+f_2(R)L_m$} (m-3-1)
					(m-2-3) edge node [left,rotate=90] {$f=f_1(T)+f_2(T)L_m$} (m-3-3)
					(m-3-3) edge node [left,rotate=90] {$f_2(T)=1$} (m-4-3)
					(m-4-3) edge node [above] {$f_1(T)=T$} (m-5-2)
						(m-4-2) edge node [right,rotate=90] {$f_1(B)=0$} (m-5-2)
							(m-4-2) edge node [left,rotate=90] {$C_1=1$} (m-5-2)
						(m-4-1) edge node [above] {$f_1(R)=R$} (m-5-2)
						(m-3-1) edge node [right,rotate=90] {$f_2(R)=1$} (m-4-1)
						(m-3-2) edge node [right,rotate=90] {$f_2(B)=1$} (m-4-2);
	\end{tikzpicture}
\caption{Relationship between different modified gravity models and General Relativity.}
\label{figtrace}
\end{figure}
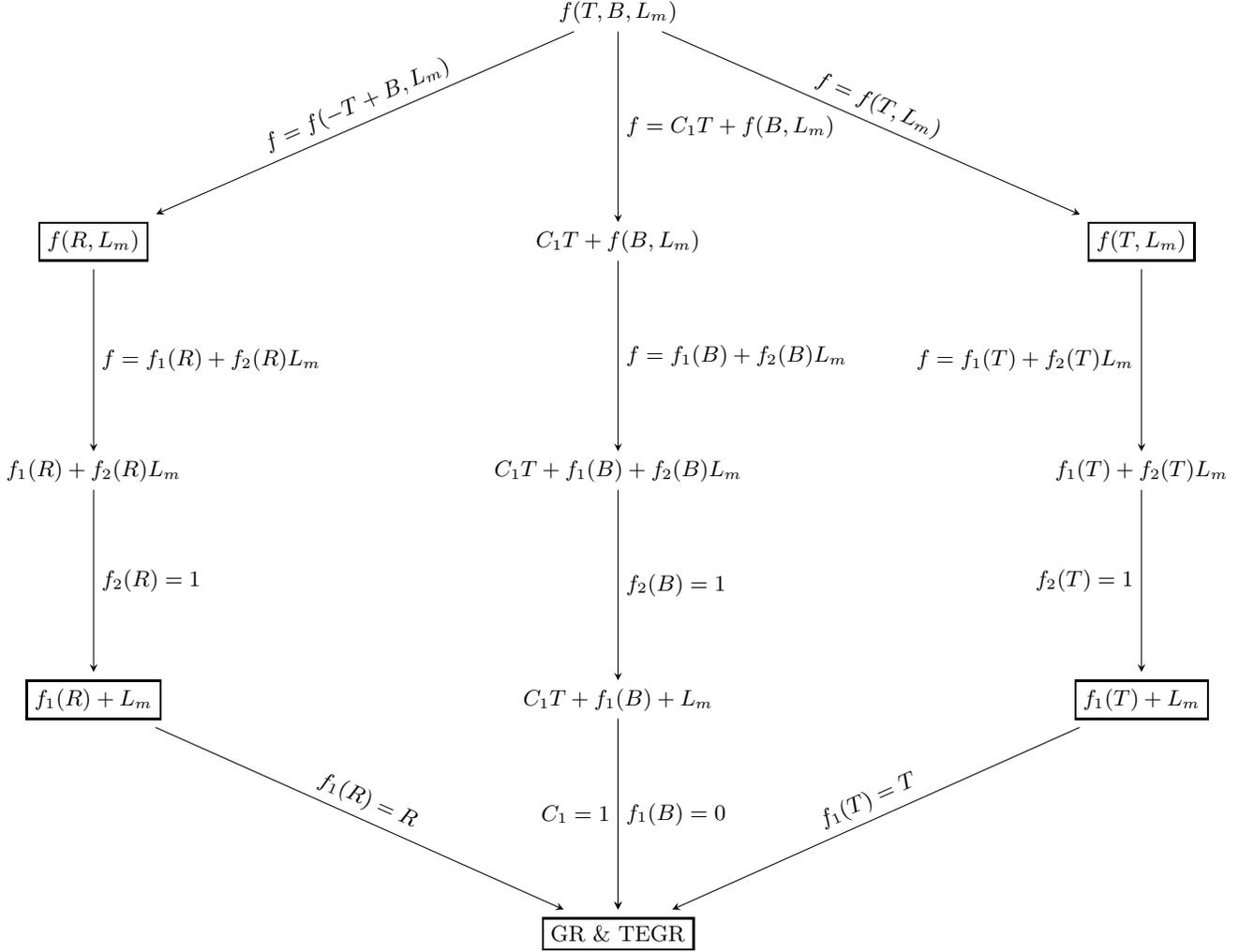

In this work we have presented a new modified theory of gravity based on an arbitrary function $f$ which depends on the scalar torsion $T$, the boundary term $B$ and the matter Lagrangian of matter $L_m$. Different kind of modified theories of gravity can be recovered from this theory. The incorporation of  $B$ in this function is with the aim to have the possibility to recover and connect standard metric theories based on the curvature scalar. This is possible since $R=-T+B$, so that it is possible to recover the generalised curvature-matter Lagrangian coupled theory $f(R,L_m)$. Fig.~\ref{figtrace} shows the most important theories that can be constructed from our action. The graph is divided into three main parts. The left part of the figure represents the scalar-curvature or standard metric theories coupled with the matter Lagrangian. Different interesting cases can be recovered from this branch, such as a generalised $f(R,L_m)$ theory or a non-minimally scalar curvature-matter coupled gravity $f_1(B)+f_2(B)L_m$ or just standard $f(R)$ gravity. The entries at the middle of the figure represent all the theories based on the boundary term $B$ and the matter Lagrangian $L_m$. In this branch, new kind of theories are presented based on a general new theory $C_1T+f(B,L_m)$, where the term $C_1 T$ is added in the model to have TEGR (or GR) in the background. The right part of the figure is related to teleparallel theories constructed by the torsion scalar and the matter Lagrangian. Under these models, a new general theory $f=f(T,L_m)$ is highlighted in box, allowing to have new kind of theories with new possible couplings between $T$ and $L_m$. As example, in this paper we have considered theories with exponential or power-law couplings between $T$ and $L_m$. Under special limits, these theories can represent a small deviation of standard TEGR with matter with or without a cosmological constant. As special case, this theory can also become a non-minimally torsion-matter coupled gravity theory $f=f_1(T)+f_2(T)L_m$, presented previously in \cite{Harko:2010mv}. Thus, different gravity curvature-matter or torsion-matter coupled theories can be constructed. Some of them have been considered  and studied in the past but others are new. The relationship between all of those well-known theories have not been established yet. From the figure, one can directly see the connection between modified teleparallel theories and standard modified theories. The quantity $B$ connects the right and left part of the figure. Hence, the connection between the teleparallel and standard theories is directly related to this boundary term $B$. Therefore, one can directly see that the mother of all of those gravity theories coupled with the matter Lagrangian is the one presented in this work, the so-called $f(T,B,L_m)$.  
%\iffalse
%\fi
In this work, we have also studied flat FLRW cosmology for the general $f(T,B,L_m)$ theory of gravity. Explicitly, we have focused our study on the dynamical systems of the full theory. In general, the theory is very complicated to work since it becomes a 10 dimensional dynamical system. This is somehow expected since the theory is very general and complicated. Using the full dynamical system found for the full theory, we then study different special interesting theories of gravity. For the case $f=f(-T+B,L_m)=f(R,L_m)$, it was proved that our full dynamical system becomes a 5-dimensional one. Moreover, we have proved how one can relate our dimensionless variables with the ones used in \cite{Azevedo:2016ehy} giving us a possibility of checking our calculations. We have found that the dynamics of this model is the same as it was described in \cite{Azevedo:2016ehy}. \\

The case $f=f(T,L_m)$ is also studied, where in general the dynamical system can be reduced to be a 3-dimensional one. This theory is analogous to $f(R,L_m)$ but mathematically speaking, it is different. It is easier to solve analytically the flat modified FLRW for a specific model for the $f(T,L_m)$ than  $f(R,L_m)$. Further, for the later theory, for the exponential/power-law curvature-matter couplings one needs to study the dynamical system to understand the dynamics. For the $f(T,L_m)$ case, the exponential/power-law torsion-matter couplings are directly integrated, giving us a scale factor of the universe directly from the modified FLRW equations. Hence, one does not need dynamical system technique to analyse the dynamics of those two examples. Another special interesting case studied was $f(T,L_m)=f_1(T)+f_2(T)L_m$. The dynamical system for this case is reduced as a 2-dimensional one. We have proved that our dimensionless variables can be directly connected to the ones introduced in \cite{Carloni:2015lsa}. This also gives us a good consistency check that our full 10-dimensional dynamical system is correct mathematically, at least for those special cases. Thus, the dynamics of those models are consistent with the study made in \cite{Carloni:2015lsa}. \\

Finally, we have also studied the dynamics of modified FLRW for $C_1T+f(B,L_m)$ gravity using dynamical system. The dynamical system for this case becomes a 5-dimensional one, exactly as the $f(R,L_m)$ case. The dynamics for this model is more complicated than $f(T,L_m)$. This is somehow expected since $B$ contains second derivatives of the scale factor and $T$ only contains first derivatives of the scale factor (see Eq.~\ref{RTB}). Further, $R$ also contains second derivatives of the scale factor, exactly as $B$, so it is not so strange to see that the dimensionality of the dynamical system of $f(R,L_m)$ is the same as $C_1T+f(B,L_m)$. Under the boundary-matter coupled model, we have studied a specific case where the matter Lagrangian is non-minimally coupled with $B$ as $f_1(B)+f_2(B)L_m$. By assuming some power-law boundary functions $f_1(B)=C_5B^{s}$ and $f_2(B)=(C_4+C_3 B^{q})$, we analysed the dynamics using dynamical system techniques. In general, the dynamical system for this power-law couplings are 4 dimensional but for the specific case where $q=1-s$, becomes a 3 dimensional one. Thus, we have analysed this model depending on three different limit cases: (i) $q=1$, (ii) $C_5=0$ and lastly the case (iii) $C_5\neq0, q=1-s$. In general, the dynamics of all of these models are similar.  As we have seen, mainly only one critical point is obtained for mainly all of them. The stability of those points were also studied,  showing the regions where the critical points become stable.\\

As a future work, it might be interesting to study further other models that can be constructed from the full theory. In principle, one can use the same 10 dynamical system that we constructed here, and then simplify it by assuming other new kind of couplings between $T$,$B$ or $L_m$. In addition, one can also use the reconstruction technique to find out which model could represent better  current cosmological observations. Further, we can also incorporate the teleparallel Gauss-Bonnet terms $T_G$ and $B_G$ to have a more general theory $f(T,B,L_m,T_G,B_G)$ (see \cite{Bahamonde:2016kba}) or even a more general new classes of theories based on the squares of the decomposition of torsion $T_{\rm ax},T_{\rm vec}$ and $T_{\rm ten}$ (see \cite{Bahamonde:2017wwk}). Then, one can study the dynamics of the modified FLRW for this general theory. By doing all of this, it will give a powerful tool to determine which models are better describing the current acceleration of the Universe, or other cosmological important questions.

\begin{acknowledgments}
The author would like to thank Christian B\"{o}hmer for his invaluable feedback and for helping to improve the manuscript. The author is supported by the Comisi{\'o}n Nacional de Investigaci{\'o}n Cient{\'{\i}}fica y Tecnol{\'o}gica (Becas Chile Grant No.~72150066).
\end{acknowledgments}

\bibliography{bibtele}

\begin{thebibliography}{56}%
\makeatletter
\providecommand \@ifxundefined [1]{%
 \@ifx{#1\undefined}
}%
\providecommand \@ifnum [1]{%
 \ifnum #1\expandafter \@firstoftwo
 \else \expandafter \@secondoftwo
 \fi
}%
\providecommand \@ifx [1]{%
 \ifx #1\expandafter \@firstoftwo
 \else \expandafter \@secondoftwo
 \fi
}%
\providecommand \natexlab [1]{#1}%
\providecommand \enquote  [1]{``#1''}%
\providecommand \bibnamefont  [1]{#1}%
\providecommand \bibfnamefont [1]{#1}%
\providecommand \citenamefont [1]{#1}%
\providecommand \href@noop [0]{\@secondoftwo}%
\providecommand \href [0]{\begingroup \@sanitize@url \@href}%
\providecommand \@href[1]{\@@startlink{#1}\@@href}%
\providecommand \@@href[1]{\endgroup#1\@@endlink}%
\providecommand \@sanitize@url [0]{\catcode `\\12\catcode `\$12\catcode
  `\&12\catcode `\#12\catcode `\^12\catcode `\_12\catcode `\%12\relax}%
\providecommand \@@startlink[1]{}%
\providecommand \@@endlink[0]{}%
\providecommand \url  [0]{\begingroup\@sanitize@url \@url }%
\providecommand \@url [1]{\endgroup\@href {#1}{\urlprefix }}%
\providecommand \urlprefix  [0]{URL }%
\providecommand \Eprint [0]{\href }%
\providecommand \doibase [0]{http://dx.doi.org/}%
\providecommand \selectlanguage [0]{\@gobble}%
\providecommand \bibinfo  [0]{\@secondoftwo}%
\providecommand \bibfield  [0]{\@secondoftwo}%
\providecommand \translation [1]{[#1]}%
\providecommand \BibitemOpen [0]{}%
\providecommand \bibitemStop [0]{}%
\providecommand \bibitemNoStop [0]{.\EOS\space}%
\providecommand \EOS [0]{\spacefactor3000\relax}%
\providecommand \BibitemShut  [1]{\csname bibitem#1\endcsname}%
\let\auto@bib@innerbib\@empty
%</preamble>
\bibitem [{\citenamefont {Riess}\ \emph {et~al.}(1998)\citenamefont {Riess},
  \citenamefont {Filippenko}, \citenamefont {Challis}, \citenamefont
  {Clocchiatti}, \citenamefont {Diercks}, \citenamefont {Garnavich},
  \citenamefont {Gilliland}, \citenamefont {Hogan}, \citenamefont {Jha},
  \citenamefont {Kirshner} \emph {et~al.}}]{Riess:1998cb}%
  \BibitemOpen
  \bibfield  {author} {\bibinfo {author} {\bibfnamefont {A.~G.}\ \bibnamefont
  {Riess}}, \bibinfo {author} {\bibfnamefont {A.~V.}\ \bibnamefont
  {Filippenko}}, \bibinfo {author} {\bibfnamefont {P.}~\bibnamefont {Challis}},
  \bibinfo {author} {\bibfnamefont {A.}~\bibnamefont {Clocchiatti}}, \bibinfo
  {author} {\bibfnamefont {A.}~\bibnamefont {Diercks}}, \bibinfo {author}
  {\bibfnamefont {P.~M.}\ \bibnamefont {Garnavich}}, \bibinfo {author}
  {\bibfnamefont {R.~L.}\ \bibnamefont {Gilliland}}, \bibinfo {author}
  {\bibfnamefont {C.~J.}\ \bibnamefont {Hogan}}, \bibinfo {author}
  {\bibfnamefont {S.}~\bibnamefont {Jha}}, \bibinfo {author} {\bibfnamefont
  {R.~P.}\ \bibnamefont {Kirshner}},  \emph {et~al.},\ }\href@noop {}
  {\bibfield  {journal} {\bibinfo  {journal} {The Astronomical Journal}\
  }\textbf {\bibinfo {volume} {116}},\ \bibinfo {pages} {1009} (\bibinfo {year}
  {1998})}\BibitemShut {NoStop}%
\bibitem [{\citenamefont {Spergel}\ \emph {et~al.}(2003)\citenamefont
  {Spergel}, \citenamefont {Verde}, \citenamefont {Peiris}, \citenamefont
  {Komatsu}, \citenamefont {Nolta}, \citenamefont {Bennett}, \citenamefont
  {Halpern}, \citenamefont {Hinshaw}, \citenamefont {Jarosik}, \citenamefont
  {Kogut} \emph {et~al.}}]{Spergel:2003cb}%
  \BibitemOpen
  \bibfield  {author} {\bibinfo {author} {\bibfnamefont {D.~N.}\ \bibnamefont
  {Spergel}}, \bibinfo {author} {\bibfnamefont {L.}~\bibnamefont {Verde}},
  \bibinfo {author} {\bibfnamefont {H.~V.}\ \bibnamefont {Peiris}}, \bibinfo
  {author} {\bibfnamefont {E.}~\bibnamefont {Komatsu}}, \bibinfo {author}
  {\bibfnamefont {M.}~\bibnamefont {Nolta}}, \bibinfo {author} {\bibfnamefont
  {C.}~\bibnamefont {Bennett}}, \bibinfo {author} {\bibfnamefont
  {M.}~\bibnamefont {Halpern}}, \bibinfo {author} {\bibfnamefont
  {G.}~\bibnamefont {Hinshaw}}, \bibinfo {author} {\bibfnamefont
  {N.}~\bibnamefont {Jarosik}}, \bibinfo {author} {\bibfnamefont
  {A.}~\bibnamefont {Kogut}},  \emph {et~al.},\ }\href@noop {} {\bibfield
  {journal} {\bibinfo  {journal} {The Astrophysical Journal Supplement Series}\
  }\textbf {\bibinfo {volume} {148}},\ \bibinfo {pages} {175} (\bibinfo {year}
  {2003})}\BibitemShut {NoStop}%
\bibitem [{\citenamefont {Spergel}\ \emph {et~al.}(2007)\citenamefont
  {Spergel}, \citenamefont {Bean}, \citenamefont {Dor{\'e}}, \citenamefont
  {Nolta}, \citenamefont {Bennett}, \citenamefont {Dunkley}, \citenamefont
  {Hinshaw}, \citenamefont {Jarosik}, \citenamefont {Komatsu}, \citenamefont
  {Page} \emph {et~al.}}]{Spergel:2006hy}%
  \BibitemOpen
  \bibfield  {author} {\bibinfo {author} {\bibfnamefont {D.~N.}\ \bibnamefont
  {Spergel}}, \bibinfo {author} {\bibfnamefont {R.}~\bibnamefont {Bean}},
  \bibinfo {author} {\bibfnamefont {O.}~\bibnamefont {Dor{\'e}}}, \bibinfo
  {author} {\bibfnamefont {M.}~\bibnamefont {Nolta}}, \bibinfo {author}
  {\bibfnamefont {C.}~\bibnamefont {Bennett}}, \bibinfo {author} {\bibfnamefont
  {J.}~\bibnamefont {Dunkley}}, \bibinfo {author} {\bibfnamefont
  {G.}~\bibnamefont {Hinshaw}}, \bibinfo {author} {\bibfnamefont {N.~e.}\
  \bibnamefont {Jarosik}}, \bibinfo {author} {\bibfnamefont {E.}~\bibnamefont
  {Komatsu}}, \bibinfo {author} {\bibfnamefont {L.}~\bibnamefont {Page}},
  \emph {et~al.},\ }\href@noop {} {\bibfield  {journal} {\bibinfo  {journal}
  {The Astrophysical Journal Supplement Series}\ }\textbf {\bibinfo {volume}
  {170}},\ \bibinfo {pages} {377} (\bibinfo {year} {2007})}\BibitemShut
  {NoStop}%
\bibitem [{\citenamefont {Komatsu}\ \emph {et~al.}(2009)\citenamefont
  {Komatsu}, \citenamefont {Dunkley}, \citenamefont {Nolta}, \citenamefont
  {Bennett}, \citenamefont {Gold}, \citenamefont {Hinshaw}, \citenamefont
  {Jarosik}, \citenamefont {Larson}, \citenamefont {Limon}, \citenamefont
  {Page} \emph {et~al.}}]{Komatsu:2008hk}%
  \BibitemOpen
  \bibfield  {author} {\bibinfo {author} {\bibfnamefont {E.}~\bibnamefont
  {Komatsu}}, \bibinfo {author} {\bibfnamefont {J.}~\bibnamefont {Dunkley}},
  \bibinfo {author} {\bibfnamefont {M.}~\bibnamefont {Nolta}}, \bibinfo
  {author} {\bibfnamefont {C.}~\bibnamefont {Bennett}}, \bibinfo {author}
  {\bibfnamefont {B.}~\bibnamefont {Gold}}, \bibinfo {author} {\bibfnamefont
  {G.}~\bibnamefont {Hinshaw}}, \bibinfo {author} {\bibfnamefont
  {N.}~\bibnamefont {Jarosik}}, \bibinfo {author} {\bibfnamefont
  {D.}~\bibnamefont {Larson}}, \bibinfo {author} {\bibfnamefont
  {M.}~\bibnamefont {Limon}}, \bibinfo {author} {\bibfnamefont
  {L.}~\bibnamefont {Page}},  \emph {et~al.},\ }\href@noop {} {\bibfield
  {journal} {\bibinfo  {journal} {The Astrophysical Journal Supplement Series}\
  }\textbf {\bibinfo {volume} {180}},\ \bibinfo {pages} {330} (\bibinfo {year}
  {2009})}\BibitemShut {NoStop}%
\bibitem [{\citenamefont {Komatsu}\ \emph {et~al.}(2011)\citenamefont
  {Komatsu}, \citenamefont {Smith}, \citenamefont {Dunkley}, \citenamefont
  {Bennett}, \citenamefont {Gold}, \citenamefont {Hinshaw}, \citenamefont
  {Jarosik}, \citenamefont {Larson}, \citenamefont {Nolta}, \citenamefont
  {Page} \emph {et~al.}}]{Komatsu:2010fb}%
  \BibitemOpen
  \bibfield  {author} {\bibinfo {author} {\bibfnamefont {E.}~\bibnamefont
  {Komatsu}}, \bibinfo {author} {\bibfnamefont {K.}~\bibnamefont {Smith}},
  \bibinfo {author} {\bibfnamefont {J.}~\bibnamefont {Dunkley}}, \bibinfo
  {author} {\bibfnamefont {C.}~\bibnamefont {Bennett}}, \bibinfo {author}
  {\bibfnamefont {B.}~\bibnamefont {Gold}}, \bibinfo {author} {\bibfnamefont
  {G.}~\bibnamefont {Hinshaw}}, \bibinfo {author} {\bibfnamefont
  {N.}~\bibnamefont {Jarosik}}, \bibinfo {author} {\bibfnamefont
  {D.}~\bibnamefont {Larson}}, \bibinfo {author} {\bibfnamefont
  {M.}~\bibnamefont {Nolta}}, \bibinfo {author} {\bibfnamefont
  {L.}~\bibnamefont {Page}},  \emph {et~al.},\ }\href@noop {} {\bibfield
  {journal} {\bibinfo  {journal} {The Astrophysical Journal Supplement Series}\
  }\textbf {\bibinfo {volume} {192}},\ \bibinfo {pages} {18} (\bibinfo {year}
  {2011})}\BibitemShut {NoStop}%
\bibitem [{\citenamefont {Eisenstein}\ \emph {et~al.}(2005)\citenamefont
  {Eisenstein}, \citenamefont {Zehavi}, \citenamefont {Hogg}, \citenamefont
  {Scoccimarro}, \citenamefont {Blanton}, \citenamefont {Nichol}, \citenamefont
  {Scranton}, \citenamefont {Seo}, \citenamefont {Tegmark}, \citenamefont
  {Zheng} \emph {et~al.}}]{Eisenstein:2005su}%
  \BibitemOpen
  \bibfield  {author} {\bibinfo {author} {\bibfnamefont {D.~J.}\ \bibnamefont
  {Eisenstein}}, \bibinfo {author} {\bibfnamefont {I.}~\bibnamefont {Zehavi}},
  \bibinfo {author} {\bibfnamefont {D.~W.}\ \bibnamefont {Hogg}}, \bibinfo
  {author} {\bibfnamefont {R.}~\bibnamefont {Scoccimarro}}, \bibinfo {author}
  {\bibfnamefont {M.~R.}\ \bibnamefont {Blanton}}, \bibinfo {author}
  {\bibfnamefont {R.~C.}\ \bibnamefont {Nichol}}, \bibinfo {author}
  {\bibfnamefont {R.}~\bibnamefont {Scranton}}, \bibinfo {author}
  {\bibfnamefont {H.-J.}\ \bibnamefont {Seo}}, \bibinfo {author} {\bibfnamefont
  {M.}~\bibnamefont {Tegmark}}, \bibinfo {author} {\bibfnamefont
  {Z.}~\bibnamefont {Zheng}},  \emph {et~al.},\ }\href@noop {} {\bibfield
  {journal} {\bibinfo  {journal} {The Astrophysical Journal}\ }\textbf
  {\bibinfo {volume} {633}},\ \bibinfo {pages} {560} (\bibinfo {year}
  {2005})}\BibitemShut {NoStop}%
\bibitem [{\citenamefont {Tegmark}\ \emph {et~al.}(2004)\citenamefont {Tegmark}
  \emph {et~al.}}]{Tegmark:2003ud}%
  \BibitemOpen
  \bibfield  {author} {\bibinfo {author} {\bibfnamefont {M.}~\bibnamefont
  {Tegmark}} \emph {et~al.} (\bibinfo {collaboration} {SDSS}),\ }\href
  {\doibase 10.1103/PhysRevD.69.103501} {\bibfield  {journal} {\bibinfo
  {journal} {Phys. Rev.}\ }\textbf {\bibinfo {volume} {D69}},\ \bibinfo {pages}
  {103501} (\bibinfo {year} {2004})},\ \Eprint
  {http://arxiv.org/abs/astro-ph/0310723} {arXiv:astro-ph/0310723 [astro-ph]}
  \BibitemShut {NoStop}%
%%CITATION = ASTRO-PH/0310723;%%
\bibitem [{\citenamefont {Nojiri}\ and\ \citenamefont
  {Odintsov}(2006)}]{Nojiri:2006ri}%
  \BibitemOpen
  \bibfield  {author} {\bibinfo {author} {\bibfnamefont {S.}~\bibnamefont
  {Nojiri}}\ and\ \bibinfo {author} {\bibfnamefont {S.~D.}\ \bibnamefont
  {Odintsov}},\ }\bibfield  {booktitle} {\emph {\bibinfo {booktitle}
  {{Theoretical physics: Current mathematical topics in gravitation and
  cosmology. Proceedings, 42nd Karpacz Winter School, Ladek, Poland, February
  6-11, 2006}}},\ }\href {\doibase 10.1142/S0219887807001928} {\bibfield
  {journal} {\bibinfo  {journal} {eConf}\ }\textbf {\bibinfo {volume}
  {C0602061}},\ \bibinfo {pages} {06} (\bibinfo {year} {2006})},\ \bibinfo
  {note} {[Int. J. Geom. Meth. Mod. Phys.4,115(2007)]},\ \Eprint
  {http://arxiv.org/abs/hep-th/0601213} {arXiv:hep-th/0601213 [hep-th]}
  \BibitemShut {NoStop}%
%%CITATION = HEP-TH/0601213;%%
\bibitem [{\citenamefont {Capozziello}\ and\ \citenamefont
  {De~Laurentis}(2011)}]{Capozziello:2011et}%
  \BibitemOpen
  \bibfield  {author} {\bibinfo {author} {\bibfnamefont {S.}~\bibnamefont
  {Capozziello}}\ and\ \bibinfo {author} {\bibfnamefont {M.}~\bibnamefont
  {De~Laurentis}},\ }\href {\doibase 10.1016/j.physrep.2011.09.003} {\bibfield
  {journal} {\bibinfo  {journal} {Phys. Rept.}\ }\textbf {\bibinfo {volume}
  {509}},\ \bibinfo {pages} {167} (\bibinfo {year} {2011})},\ \Eprint
  {http://arxiv.org/abs/1108.6266} {arXiv:1108.6266 [gr-qc]} \BibitemShut
  {NoStop}%
%%CITATION = ARXIV:1108.6266;%%
\bibitem [{\citenamefont
  {Weitzenb{\"o}ck}(1923)}]{weitzenbock1923invariantentheorie}%
  \BibitemOpen
  \bibfield  {author} {\bibinfo {author} {\bibfnamefont {R.}~\bibnamefont
  {Weitzenb{\"o}ck}},\ }\href@noop {} {\bibfield  {journal} {\bibinfo
  {journal} {Invarianten Theorie. Nordhoff, Groningen}\ } (\bibinfo {year}
  {1923})}\BibitemShut {NoStop}%
\bibitem [{\citenamefont {Hayashi}(1977)}]{Hayashi:1977jd}%
  \BibitemOpen
  \bibfield  {author} {\bibinfo {author} {\bibfnamefont {K.}~\bibnamefont
  {Hayashi}},\ }\href {\doibase 10.1016/0370-2693(77)90840-1} {\bibfield
  {journal} {\bibinfo  {journal} {Phys. Lett.}\ }\textbf {\bibinfo {volume}
  {B69}},\ \bibinfo {pages} {441} (\bibinfo {year} {1977})}\BibitemShut
  {NoStop}%
%%CITATION = PHLTA,B69,441;%%
\bibitem [{\citenamefont {de~Andrade}\ and\ \citenamefont
  {Pereira}(1997)}]{deAndrade:1997qt}%
  \BibitemOpen
  \bibfield  {author} {\bibinfo {author} {\bibfnamefont {V.~C.}\ \bibnamefont
  {de~Andrade}}\ and\ \bibinfo {author} {\bibfnamefont {J.~G.}\ \bibnamefont
  {Pereira}},\ }\href {\doibase 10.1103/PhysRevD.56.4689} {\bibfield  {journal}
  {\bibinfo  {journal} {Phys. Rev.}\ }\textbf {\bibinfo {volume} {D56}},\
  \bibinfo {pages} {4689} (\bibinfo {year} {1997})},\ \Eprint
  {http://arxiv.org/abs/gr-qc/9703059} {arXiv:gr-qc/9703059 [gr-qc]}
  \BibitemShut {NoStop}%
%%CITATION = GR-QC/9703059;%%
\bibitem [{\citenamefont {de~Andrade}\ \emph {et~al.}(2000)\citenamefont
  {de~Andrade}, \citenamefont {Guillen},\ and\ \citenamefont
  {Pereira}}]{deAndrade:2000kr}%
  \BibitemOpen
  \bibfield  {author} {\bibinfo {author} {\bibfnamefont {V.~C.}\ \bibnamefont
  {de~Andrade}}, \bibinfo {author} {\bibfnamefont {L.~C.~T.}\ \bibnamefont
  {Guillen}}, \ and\ \bibinfo {author} {\bibfnamefont {J.~G.}\ \bibnamefont
  {Pereira}},\ }\href {\doibase 10.1103/PhysRevLett.84.4533} {\bibfield
  {journal} {\bibinfo  {journal} {Phys. Rev. Lett.}\ }\textbf {\bibinfo
  {volume} {84}},\ \bibinfo {pages} {4533} (\bibinfo {year} {2000})},\ \Eprint
  {http://arxiv.org/abs/gr-qc/0003100} {arXiv:gr-qc/0003100 [gr-qc]}
  \BibitemShut {NoStop}%
%%CITATION = GR-QC/0003100;%%
\bibitem [{\citenamefont {Arcos}\ and\ \citenamefont
  {Pereira}(2004)}]{Arcos:2005ec}%
  \BibitemOpen
  \bibfield  {author} {\bibinfo {author} {\bibfnamefont {H.~I.}\ \bibnamefont
  {Arcos}}\ and\ \bibinfo {author} {\bibfnamefont {J.~G.}\ \bibnamefont
  {Pereira}},\ }\href {\doibase 10.1142/S0218271804006462} {\bibfield
  {journal} {\bibinfo  {journal} {Int. J. Mod. Phys.}\ }\textbf {\bibinfo
  {volume} {D13}},\ \bibinfo {pages} {2193} (\bibinfo {year} {2004})},\ \Eprint
  {http://arxiv.org/abs/gr-qc/0501017} {arXiv:gr-qc/0501017 [gr-qc]}
  \BibitemShut {NoStop}%
%%CITATION = GR-QC/0501017;%%
\bibitem [{\citenamefont {Obukhov}\ and\ \citenamefont
  {Pereira}(2003)}]{Obukhov:2002tm}%
  \BibitemOpen
  \bibfield  {author} {\bibinfo {author} {\bibfnamefont {{\relax Yu}.~N.}\
  \bibnamefont {Obukhov}}\ and\ \bibinfo {author} {\bibfnamefont {J.~G.}\
  \bibnamefont {Pereira}},\ }\href {\doibase 10.1103/PhysRevD.67.044016}
  {\bibfield  {journal} {\bibinfo  {journal} {Phys. Rev.}\ }\textbf {\bibinfo
  {volume} {D67}},\ \bibinfo {pages} {044016} (\bibinfo {year} {2003})},\
  \Eprint {http://arxiv.org/abs/gr-qc/0212080} {arXiv:gr-qc/0212080 [gr-qc]}
  \BibitemShut {NoStop}%
%%CITATION = GR-QC/0212080;%%
\bibitem [{\citenamefont {Maluf}(2013)}]{Maluf:2013gaa}%
  \BibitemOpen
  \bibfield  {author} {\bibinfo {author} {\bibfnamefont {J.~W.}\ \bibnamefont
  {Maluf}},\ }\href {\doibase 10.1002/andp.201200272} {\bibfield  {journal}
  {\bibinfo  {journal} {Annalen Phys.}\ }\textbf {\bibinfo {volume} {525}},\
  \bibinfo {pages} {339} (\bibinfo {year} {2013})},\ \Eprint
  {http://arxiv.org/abs/1303.3897} {arXiv:1303.3897 [gr-qc]} \BibitemShut
  {NoStop}%
%%CITATION = ARXIV:1303.3897;%%
\bibitem [{\citenamefont {Aldrovandi}\ and\ \citenamefont
  {Pereira}(2012)}]{AP}%
  \BibitemOpen
  \bibfield  {author} {\bibinfo {author} {\bibfnamefont {R.}~\bibnamefont
  {Aldrovandi}}\ and\ \bibinfo {author} {\bibfnamefont {J.~G.}\ \bibnamefont
  {Pereira}},\ }\href@noop {} {\emph {\bibinfo {title} {{Teleparallel Gravity:
  An Introduction}}}}\ (\bibinfo  {publisher} {Springer, Dordrechts},\ \bibinfo
  {year} {2012})\BibitemShut {NoStop}%
%%CITATION = INSPIRE-1235601;%%
\bibitem [{\citenamefont {Bengochea}\ and\ \citenamefont
  {Ferraro}(2009)}]{Bengochea:2008gz}%
  \BibitemOpen
  \bibfield  {author} {\bibinfo {author} {\bibfnamefont {G.~R.}\ \bibnamefont
  {Bengochea}}\ and\ \bibinfo {author} {\bibfnamefont {R.}~\bibnamefont
  {Ferraro}},\ }\href {\doibase 10.1103/PhysRevD.79.124019} {\bibfield
  {journal} {\bibinfo  {journal} {Phys. Rev.}\ }\textbf {\bibinfo {volume}
  {D79}},\ \bibinfo {pages} {124019} (\bibinfo {year} {2009})},\ \Eprint
  {http://arxiv.org/abs/0812.1205} {arXiv:0812.1205 [astro-ph]} \BibitemShut
  {NoStop}%
%%CITATION = ARXIV:0812.1205;%%
\bibitem [{\citenamefont {Ferraro}\ and\ \citenamefont
  {Fiorini}(2007)}]{Ferraro:2006jd}%
  \BibitemOpen
  \bibfield  {author} {\bibinfo {author} {\bibfnamefont {R.}~\bibnamefont
  {Ferraro}}\ and\ \bibinfo {author} {\bibfnamefont {F.}~\bibnamefont
  {Fiorini}},\ }\href {\doibase 10.1103/PhysRevD.75.084031} {\bibfield
  {journal} {\bibinfo  {journal} {Phys. Rev.}\ }\textbf {\bibinfo {volume}
  {D75}},\ \bibinfo {pages} {084031} (\bibinfo {year} {2007})},\ \Eprint
  {http://arxiv.org/abs/gr-qc/0610067} {arXiv:gr-qc/0610067 [gr-qc]}
  \BibitemShut {NoStop}%
%%CITATION = GR-QC/0610067;%%
\bibitem [{\citenamefont {Bengochea}(2011)}]{Bengochea:2010sg}%
  \BibitemOpen
  \bibfield  {author} {\bibinfo {author} {\bibfnamefont {G.~R.}\ \bibnamefont
  {Bengochea}},\ }\href {\doibase 10.1016/j.physletb.2010.11.064} {\bibfield
  {journal} {\bibinfo  {journal} {Phys. Lett.}\ }\textbf {\bibinfo {volume}
  {B695}},\ \bibinfo {pages} {405} (\bibinfo {year} {2011})},\ \Eprint
  {http://arxiv.org/abs/1008.3188} {arXiv:1008.3188 [astro-ph.CO]} \BibitemShut
  {NoStop}%
%%CITATION = ARXIV:1008.3188;%%
\bibitem [{\citenamefont {Li}\ \emph {et~al.}(2011{\natexlab{a}})\citenamefont
  {Li}, \citenamefont {Sotiriou},\ and\ \citenamefont {Barrow}}]{Li:2011wu}%
  \BibitemOpen
  \bibfield  {author} {\bibinfo {author} {\bibfnamefont {B.}~\bibnamefont
  {Li}}, \bibinfo {author} {\bibfnamefont {T.~P.}\ \bibnamefont {Sotiriou}}, \
  and\ \bibinfo {author} {\bibfnamefont {J.~D.}\ \bibnamefont {Barrow}},\
  }\href {\doibase 10.1103/PhysRevD.83.104017} {\bibfield  {journal} {\bibinfo
  {journal} {Phys. Rev.}\ }\textbf {\bibinfo {volume} {D83}},\ \bibinfo {pages}
  {104017} (\bibinfo {year} {2011}{\natexlab{a}})},\ \Eprint
  {http://arxiv.org/abs/1103.2786} {arXiv:1103.2786 [astro-ph.CO]} \BibitemShut
  {NoStop}%
%%CITATION = ARXIV:1103.2786;%%
\bibitem [{\citenamefont {Wu}\ and\ \citenamefont {Yu}(2011)}]{Wu:2010av}%
  \BibitemOpen
  \bibfield  {author} {\bibinfo {author} {\bibfnamefont {P.}~\bibnamefont
  {Wu}}\ and\ \bibinfo {author} {\bibfnamefont {H.~W.}\ \bibnamefont {Yu}},\
  }\href {\doibase 10.1140/epjc/s10052-011-1552-2} {\bibfield  {journal}
  {\bibinfo  {journal} {Eur. Phys. J.}\ }\textbf {\bibinfo {volume} {C71}},\
  \bibinfo {pages} {1552} (\bibinfo {year} {2011})},\ \Eprint
  {http://arxiv.org/abs/1008.3669} {arXiv:1008.3669 [gr-qc]} \BibitemShut
  {NoStop}%
%%CITATION = ARXIV:1008.3669;%%
\bibitem [{\citenamefont {Wu}\ and\ \citenamefont
  {Yu}(2010{\natexlab{a}})}]{Wu:2010mn}%
  \BibitemOpen
  \bibfield  {author} {\bibinfo {author} {\bibfnamefont {P.}~\bibnamefont
  {Wu}}\ and\ \bibinfo {author} {\bibfnamefont {H.~W.}\ \bibnamefont {Yu}},\
  }\href {\doibase 10.1016/j.physletb.2010.08.073} {\bibfield  {journal}
  {\bibinfo  {journal} {Phys. Lett.}\ }\textbf {\bibinfo {volume} {B693}},\
  \bibinfo {pages} {415} (\bibinfo {year} {2010}{\natexlab{a}})},\ \Eprint
  {http://arxiv.org/abs/1006.0674} {arXiv:1006.0674 [gr-qc]} \BibitemShut
  {NoStop}%
%%CITATION = ARXIV:1006.0674;%%
\bibitem [{\citenamefont {Dent}\ \emph {et~al.}(2011)\citenamefont {Dent},
  \citenamefont {Dutta},\ and\ \citenamefont {Saridakis}}]{Dent:2011zz}%
  \BibitemOpen
  \bibfield  {author} {\bibinfo {author} {\bibfnamefont {J.~B.}\ \bibnamefont
  {Dent}}, \bibinfo {author} {\bibfnamefont {S.}~\bibnamefont {Dutta}}, \ and\
  \bibinfo {author} {\bibfnamefont {E.~N.}\ \bibnamefont {Saridakis}},\ }\href
  {\doibase 10.1088/1475-7516/2011/01/009} {\bibfield  {journal} {\bibinfo
  {journal} {JCAP}\ }\textbf {\bibinfo {volume} {1101}},\ \bibinfo {pages}
  {009} (\bibinfo {year} {2011})},\ \Eprint {http://arxiv.org/abs/1010.2215}
  {arXiv:1010.2215 [astro-ph.CO]} \BibitemShut {NoStop}%
%%CITATION = ARXIV:1010.2215;%%
\bibitem [{\citenamefont {Chen}\ \emph {et~al.}(2011)\citenamefont {Chen},
  \citenamefont {Dent}, \citenamefont {Dutta},\ and\ \citenamefont
  {Saridakis}}]{Chen:2010va}%
  \BibitemOpen
  \bibfield  {author} {\bibinfo {author} {\bibfnamefont {S.-H.}\ \bibnamefont
  {Chen}}, \bibinfo {author} {\bibfnamefont {J.~B.}\ \bibnamefont {Dent}},
  \bibinfo {author} {\bibfnamefont {S.}~\bibnamefont {Dutta}}, \ and\ \bibinfo
  {author} {\bibfnamefont {E.~N.}\ \bibnamefont {Saridakis}},\ }\href {\doibase
  10.1103/PhysRevD.83.023508} {\bibfield  {journal} {\bibinfo  {journal} {Phys.
  Rev.}\ }\textbf {\bibinfo {volume} {D83}},\ \bibinfo {pages} {023508}
  (\bibinfo {year} {2011})},\ \Eprint {http://arxiv.org/abs/1008.1250}
  {arXiv:1008.1250 [astro-ph.CO]} \BibitemShut {NoStop}%
%%CITATION = ARXIV:1008.1250;%%
\bibitem [{\citenamefont {Cai}\ \emph {et~al.}(2011)\citenamefont {Cai},
  \citenamefont {Chen}, \citenamefont {Dent}, \citenamefont {Dutta},\ and\
  \citenamefont {Saridakis}}]{Cai:2011tc}%
  \BibitemOpen
  \bibfield  {author} {\bibinfo {author} {\bibfnamefont {Y.-F.}\ \bibnamefont
  {Cai}}, \bibinfo {author} {\bibfnamefont {S.-H.}\ \bibnamefont {Chen}},
  \bibinfo {author} {\bibfnamefont {J.~B.}\ \bibnamefont {Dent}}, \bibinfo
  {author} {\bibfnamefont {S.}~\bibnamefont {Dutta}}, \ and\ \bibinfo {author}
  {\bibfnamefont {E.~N.}\ \bibnamefont {Saridakis}},\ }\href {\doibase
  10.1088/0264-9381/28/21/215011} {\bibfield  {journal} {\bibinfo  {journal}
  {Class. Quant. Grav.}\ }\textbf {\bibinfo {volume} {28}},\ \bibinfo {pages}
  {215011} (\bibinfo {year} {2011})},\ \Eprint {http://arxiv.org/abs/1104.4349}
  {arXiv:1104.4349 [astro-ph.CO]} \BibitemShut {NoStop}%
%%CITATION = ARXIV:1104.4349;%%
\bibitem [{\citenamefont {Bamba}\ \emph {et~al.}(2011)\citenamefont {Bamba},
  \citenamefont {Geng}, \citenamefont {Lee},\ and\ \citenamefont
  {Luo}}]{Bamba:2010wb}%
  \BibitemOpen
  \bibfield  {author} {\bibinfo {author} {\bibfnamefont {K.}~\bibnamefont
  {Bamba}}, \bibinfo {author} {\bibfnamefont {C.-Q.}\ \bibnamefont {Geng}},
  \bibinfo {author} {\bibfnamefont {C.-C.}\ \bibnamefont {Lee}}, \ and\
  \bibinfo {author} {\bibfnamefont {L.-W.}\ \bibnamefont {Luo}},\ }\href
  {\doibase 10.1088/1475-7516/2011/01/021} {\bibfield  {journal} {\bibinfo
  {journal} {JCAP}\ }\textbf {\bibinfo {volume} {1101}},\ \bibinfo {pages}
  {021} (\bibinfo {year} {2011})},\ \Eprint {http://arxiv.org/abs/1011.0508}
  {arXiv:1011.0508 [astro-ph.CO]} \BibitemShut {NoStop}%
%%CITATION = ARXIV:1011.0508;%%
\bibitem [{\citenamefont {Biswas}\ and\ \citenamefont
  {Chakraborty}(2015)}]{Biswas:2015cva}%
  \BibitemOpen
  \bibfield  {author} {\bibinfo {author} {\bibfnamefont {S.~K.}\ \bibnamefont
  {Biswas}}\ and\ \bibinfo {author} {\bibfnamefont {S.}~\bibnamefont
  {Chakraborty}},\ }\href {\doibase 10.1142/S0218271815500467} {\bibfield
  {journal} {\bibinfo  {journal} {Int. J. Mod. Phys.}\ }\textbf {\bibinfo
  {volume} {D24}},\ \bibinfo {pages} {1550046} (\bibinfo {year} {2015})},\
  \Eprint {http://arxiv.org/abs/1504.02431} {arXiv:1504.02431 [gr-qc]}
  \BibitemShut {NoStop}%
%%CITATION = ARXIV:1504.02431;%%
\bibitem [{\citenamefont {Wu}\ and\ \citenamefont
  {Yu}(2010{\natexlab{b}})}]{Wu:2010xk}%
  \BibitemOpen
  \bibfield  {author} {\bibinfo {author} {\bibfnamefont {P.}~\bibnamefont
  {Wu}}\ and\ \bibinfo {author} {\bibfnamefont {H.~W.}\ \bibnamefont {Yu}},\
  }\href {\doibase 10.1016/j.physletb.2010.07.038} {\bibfield  {journal}
  {\bibinfo  {journal} {Phys. Lett.}\ }\textbf {\bibinfo {volume} {B692}},\
  \bibinfo {pages} {176} (\bibinfo {year} {2010}{\natexlab{b}})},\ \Eprint
  {http://arxiv.org/abs/1007.2348} {arXiv:1007.2348 [astro-ph.CO]} \BibitemShut
  {NoStop}%
%%CITATION = ARXIV:1007.2348;%%
\bibitem [{\citenamefont {Nunes}\ \emph {et~al.}(2016)\citenamefont {Nunes},
  \citenamefont {Pan},\ and\ \citenamefont {Saridakis}}]{Nunes:2016qyp}%
  \BibitemOpen
  \bibfield  {author} {\bibinfo {author} {\bibfnamefont {R.~C.}\ \bibnamefont
  {Nunes}}, \bibinfo {author} {\bibfnamefont {S.}~\bibnamefont {Pan}}, \ and\
  \bibinfo {author} {\bibfnamefont {E.~N.}\ \bibnamefont {Saridakis}},\ }\href
  {\doibase 10.1088/1475-7516/2016/08/011} {\bibfield  {journal} {\bibinfo
  {journal} {JCAP}\ }\textbf {\bibinfo {volume} {1608}},\ \bibinfo {pages}
  {011} (\bibinfo {year} {2016})},\ \Eprint {http://arxiv.org/abs/1606.04359}
  {arXiv:1606.04359 [gr-qc]} \BibitemShut {NoStop}%
%%CITATION = ARXIV:1606.04359;%%
\bibitem [{\citenamefont {Zubair}\ and\ \citenamefont
  {Waheed}(2015)}]{Zubair:2014xsa}%
  \BibitemOpen
  \bibfield  {author} {\bibinfo {author} {\bibfnamefont {M.}~\bibnamefont
  {Zubair}}\ and\ \bibinfo {author} {\bibfnamefont {S.}~\bibnamefont
  {Waheed}},\ }\href {\doibase 10.1007/s10509-014-2181-7} {\bibfield  {journal}
  {\bibinfo  {journal} {Astrophys. Space Sci.}\ }\textbf {\bibinfo {volume}
  {355}},\ \bibinfo {pages} {361} (\bibinfo {year} {2015})},\ \Eprint
  {http://arxiv.org/abs/1502.03002} {arXiv:1502.03002 [gr-qc]} \BibitemShut
  {NoStop}%
%%CITATION = ARXIV:1502.03002;%%
\bibitem [{\citenamefont {Hohmann}\ \emph {et~al.}(2017)\citenamefont
  {Hohmann}, \citenamefont {Jarv},\ and\ \citenamefont
  {Ualikhanova}}]{Hohmann:2017jao}%
  \BibitemOpen
  \bibfield  {author} {\bibinfo {author} {\bibfnamefont {M.}~\bibnamefont
  {Hohmann}}, \bibinfo {author} {\bibfnamefont {L.}~\bibnamefont {Jarv}}, \
  and\ \bibinfo {author} {\bibfnamefont {U.}~\bibnamefont {Ualikhanova}},\
  }\href {\doibase 10.1103/PhysRevD.96.043508} {\bibfield  {journal} {\bibinfo
  {journal} {Phys. Rev.}\ }\textbf {\bibinfo {volume} {D96}},\ \bibinfo {pages}
  {043508} (\bibinfo {year} {2017})},\ \Eprint
  {http://arxiv.org/abs/1706.02376} {arXiv:1706.02376 [gr-qc]} \BibitemShut
  {NoStop}%
%%CITATION = ARXIV:1706.02376;%%
\bibitem [{\citenamefont {Cai}\ \emph {et~al.}(2016)\citenamefont {Cai},
  \citenamefont {Capozziello}, \citenamefont {De~Laurentis},\ and\
  \citenamefont {Saridakis}}]{Cai:2015emx}%
  \BibitemOpen
  \bibfield  {author} {\bibinfo {author} {\bibfnamefont {Y.-F.}\ \bibnamefont
  {Cai}}, \bibinfo {author} {\bibfnamefont {S.}~\bibnamefont {Capozziello}},
  \bibinfo {author} {\bibfnamefont {M.}~\bibnamefont {De~Laurentis}}, \ and\
  \bibinfo {author} {\bibfnamefont {E.~N.}\ \bibnamefont {Saridakis}},\ }\href
  {\doibase 10.1088/0034-4885/79/10/106901} {\bibfield  {journal} {\bibinfo
  {journal} {Rept. Prog. Phys.}\ }\textbf {\bibinfo {volume} {79}},\ \bibinfo
  {pages} {106901} (\bibinfo {year} {2016})},\ \Eprint
  {http://arxiv.org/abs/1511.07586} {arXiv:1511.07586 [gr-qc]} \BibitemShut
  {NoStop}%
%%CITATION = ARXIV:1511.07586;%%
\bibitem [{\citenamefont {Bahamonde}\ \emph {et~al.}(2015)\citenamefont
  {Bahamonde}, \citenamefont {{B\"ohmer}},\ and\ \citenamefont
  {Wright}}]{Bahamonde:2015zma}%
  \BibitemOpen
  \bibfield  {author} {\bibinfo {author} {\bibfnamefont {S.}~\bibnamefont
  {Bahamonde}}, \bibinfo {author} {\bibfnamefont {C.~G.}\ \bibnamefont
  {{B\"ohmer}}}, \ and\ \bibinfo {author} {\bibfnamefont {M.}~\bibnamefont
  {Wright}},\ }\href {\doibase 10.1103/PhysRevD.92.104042} {\bibfield
  {journal} {\bibinfo  {journal} {Phys. Rev.}\ }\textbf {\bibinfo {volume}
  {D92}},\ \bibinfo {pages} {104042} (\bibinfo {year} {2015})},\ \Eprint
  {http://arxiv.org/abs/1508.05120} {arXiv:1508.05120 [gr-qc]} \BibitemShut
  {NoStop}%
%%CITATION = ARXIV:1508.05120;%%
\bibitem [{\citenamefont {Bahamonde}\ \emph
  {et~al.}(2018{\natexlab{a}})\citenamefont {Bahamonde}, \citenamefont
  {Zubair},\ and\ \citenamefont {Abbas}}]{Bahamonde:2016cul}%
  \BibitemOpen
  \bibfield  {author} {\bibinfo {author} {\bibfnamefont {S.}~\bibnamefont
  {Bahamonde}}, \bibinfo {author} {\bibfnamefont {M.}~\bibnamefont {Zubair}}, \
  and\ \bibinfo {author} {\bibfnamefont {G.}~\bibnamefont {Abbas}},\ }\href
  {\doibase 10.1016/j.dark.2017.12.005} {\bibfield  {journal} {\bibinfo
  {journal} {Phys. Dark Univ.}\ }\textbf {\bibinfo {volume} {19}},\ \bibinfo
  {pages} {78} (\bibinfo {year} {2018}{\natexlab{a}})},\ \Eprint
  {http://arxiv.org/abs/1609.08373} {arXiv:1609.08373 [gr-qc]} \BibitemShut
  {NoStop}%
%%CITATION = ARXIV:1609.08373;%%
\bibitem [{\citenamefont {Bahamonde}\ and\ \citenamefont
  {Capozziello}(2017)}]{Bahamonde:2016grb}%
  \BibitemOpen
  \bibfield  {author} {\bibinfo {author} {\bibfnamefont {S.}~\bibnamefont
  {Bahamonde}}\ and\ \bibinfo {author} {\bibfnamefont {S.}~\bibnamefont
  {Capozziello}},\ }\href {\doibase 10.1140/epjc/s10052-017-4677-0} {\bibfield
  {journal} {\bibinfo  {journal} {Eur. Phys. J.}\ }\textbf {\bibinfo {volume}
  {C77}},\ \bibinfo {pages} {107} (\bibinfo {year} {2017})},\ \Eprint
  {http://arxiv.org/abs/1612.01299} {arXiv:1612.01299 [gr-qc]} \BibitemShut
  {NoStop}%
%%CITATION = ARXIV:1612.01299;%%
\bibitem [{\citenamefont {Harko}\ \emph {et~al.}(2011)\citenamefont {Harko},
  \citenamefont {Lobo}, \citenamefont {Nojiri},\ and\ \citenamefont
  {Odintsov}}]{Harko:2011kv}%
  \BibitemOpen
  \bibfield  {author} {\bibinfo {author} {\bibfnamefont {T.}~\bibnamefont
  {Harko}}, \bibinfo {author} {\bibfnamefont {F.~S.~N.}\ \bibnamefont {Lobo}},
  \bibinfo {author} {\bibfnamefont {S.}~\bibnamefont {Nojiri}}, \ and\ \bibinfo
  {author} {\bibfnamefont {S.~D.}\ \bibnamefont {Odintsov}},\ }\href {\doibase
  10.1103/PhysRevD.84.024020} {\bibfield  {journal} {\bibinfo  {journal} {Phys.
  Rev.}\ }\textbf {\bibinfo {volume} {D84}},\ \bibinfo {pages} {024020}
  (\bibinfo {year} {2011})},\ \Eprint {http://arxiv.org/abs/1104.2669}
  {arXiv:1104.2669 [gr-qc]} \BibitemShut {NoStop}%
%%CITATION = ARXIV:1104.2669;%%
\bibitem [{\citenamefont {Bertolami}\ \emph {et~al.}(2007)\citenamefont
  {Bertolami}, \citenamefont {Boehmer}, \citenamefont {Harko},\ and\
  \citenamefont {Lobo}}]{Bertolami:2007gv}%
  \BibitemOpen
  \bibfield  {author} {\bibinfo {author} {\bibfnamefont {O.}~\bibnamefont
  {Bertolami}}, \bibinfo {author} {\bibfnamefont {C.~G.}\ \bibnamefont
  {Boehmer}}, \bibinfo {author} {\bibfnamefont {T.}~\bibnamefont {Harko}}, \
  and\ \bibinfo {author} {\bibfnamefont {F.~S.~N.}\ \bibnamefont {Lobo}},\
  }\href {\doibase 10.1103/PhysRevD.75.104016} {\bibfield  {journal} {\bibinfo
  {journal} {Phys. Rev.}\ }\textbf {\bibinfo {volume} {D75}},\ \bibinfo {pages}
  {104016} (\bibinfo {year} {2007})},\ \Eprint {http://arxiv.org/abs/0704.1733}
  {arXiv:0704.1733 [gr-qc]} \BibitemShut {NoStop}%
%%CITATION = ARXIV:0704.1733;%%
\bibitem [{\citenamefont {Harko}\ \emph
  {et~al.}(2014{\natexlab{a}})\citenamefont {Harko}, \citenamefont {Lobo},
  \citenamefont {Otalora},\ and\ \citenamefont {Saridakis}}]{Harko:2014aja}%
  \BibitemOpen
  \bibfield  {author} {\bibinfo {author} {\bibfnamefont {T.}~\bibnamefont
  {Harko}}, \bibinfo {author} {\bibfnamefont {F.~S.~N.}\ \bibnamefont {Lobo}},
  \bibinfo {author} {\bibfnamefont {G.}~\bibnamefont {Otalora}}, \ and\
  \bibinfo {author} {\bibfnamefont {E.~N.}\ \bibnamefont {Saridakis}},\ }\href
  {\doibase 10.1088/1475-7516/2014/12/021} {\bibfield  {journal} {\bibinfo
  {journal} {JCAP}\ }\textbf {\bibinfo {volume} {1412}},\ \bibinfo {pages}
  {021} (\bibinfo {year} {2014}{\natexlab{a}})},\ \Eprint
  {http://arxiv.org/abs/1405.0519} {arXiv:1405.0519 [gr-qc]} \BibitemShut
  {NoStop}%
%%CITATION = ARXIV:1405.0519;%%
\bibitem [{\citenamefont {Harko}\ \emph
  {et~al.}(2014{\natexlab{b}})\citenamefont {Harko}, \citenamefont {Lobo},
  \citenamefont {Otalora},\ and\ \citenamefont {Saridakis}}]{Harko:2014sja}%
  \BibitemOpen
  \bibfield  {author} {\bibinfo {author} {\bibfnamefont {T.}~\bibnamefont
  {Harko}}, \bibinfo {author} {\bibfnamefont {F.~S.~N.}\ \bibnamefont {Lobo}},
  \bibinfo {author} {\bibfnamefont {G.}~\bibnamefont {Otalora}}, \ and\
  \bibinfo {author} {\bibfnamefont {E.~N.}\ \bibnamefont {Saridakis}},\ }\href
  {\doibase 10.1103/PhysRevD.89.124036} {\bibfield  {journal} {\bibinfo
  {journal} {Phys. Rev.}\ }\textbf {\bibinfo {volume} {D89}},\ \bibinfo {pages}
  {124036} (\bibinfo {year} {2014}{\natexlab{b}})},\ \Eprint
  {http://arxiv.org/abs/1404.6212} {arXiv:1404.6212 [gr-qc]} \BibitemShut
  {NoStop}%
%%CITATION = ARXIV:1404.6212;%%
\bibitem [{\citenamefont {Bertolami}\ \emph {et~al.}(2008)\citenamefont
  {Bertolami}, \citenamefont {Lobo},\ and\ \citenamefont
  {Paramos}}]{Bertolami:2008ab}%
  \BibitemOpen
  \bibfield  {author} {\bibinfo {author} {\bibfnamefont {O.}~\bibnamefont
  {Bertolami}}, \bibinfo {author} {\bibfnamefont {F.~S.~N.}\ \bibnamefont
  {Lobo}}, \ and\ \bibinfo {author} {\bibfnamefont {J.}~\bibnamefont
  {Paramos}},\ }\href {\doibase 10.1103/PhysRevD.78.064036} {\bibfield
  {journal} {\bibinfo  {journal} {Phys. Rev.}\ }\textbf {\bibinfo {volume}
  {D78}},\ \bibinfo {pages} {064036} (\bibinfo {year} {2008})},\ \Eprint
  {http://arxiv.org/abs/0806.4434} {arXiv:0806.4434 [gr-qc]} \BibitemShut
  {NoStop}%
%%CITATION = ARXIV:0806.4434;%%
\bibitem [{\citenamefont {Azevedo}\ and\ \citenamefont
  {Páramos}(2016)}]{Azevedo:2016ehy}%
  \BibitemOpen
  \bibfield  {author} {\bibinfo {author} {\bibfnamefont {R.~P.~L.}\
  \bibnamefont {Azevedo}}\ and\ \bibinfo {author} {\bibfnamefont
  {J.}~\bibnamefont {Páramos}},\ }\href {\doibase 10.1103/PhysRevD.94.064036}
  {\bibfield  {journal} {\bibinfo  {journal} {Phys. Rev.}\ }\textbf {\bibinfo
  {volume} {D94}},\ \bibinfo {pages} {064036} (\bibinfo {year} {2016})},\
  \Eprint {http://arxiv.org/abs/1606.08919} {arXiv:1606.08919 [gr-qc]}
  \BibitemShut {NoStop}%
%%CITATION = ARXIV:1606.08919;%%
\bibitem [{\citenamefont {An}\ \emph {et~al.}(2016)\citenamefont {An},
  \citenamefont {Xu}, \citenamefont {Wang},\ and\ \citenamefont
  {Gong}}]{An:2015mvw}%
  \BibitemOpen
  \bibfield  {author} {\bibinfo {author} {\bibfnamefont {R.}~\bibnamefont
  {An}}, \bibinfo {author} {\bibfnamefont {X.}~\bibnamefont {Xu}}, \bibinfo
  {author} {\bibfnamefont {B.}~\bibnamefont {Wang}}, \ and\ \bibinfo {author}
  {\bibfnamefont {Y.}~\bibnamefont {Gong}},\ }\href {\doibase
  10.1103/PhysRevD.93.103505} {\bibfield  {journal} {\bibinfo  {journal} {Phys.
  Rev.}\ }\textbf {\bibinfo {volume} {D93}},\ \bibinfo {pages} {103505}
  (\bibinfo {year} {2016})},\ \Eprint {http://arxiv.org/abs/1512.09281}
  {arXiv:1512.09281 [gr-qc]} \BibitemShut {NoStop}%
%%CITATION = ARXIV:1512.09281;%%
\bibitem [{\citenamefont {Carloni}\ \emph {et~al.}(2016)\citenamefont
  {Carloni}, \citenamefont {Lobo}, \citenamefont {Otalora},\ and\ \citenamefont
  {Saridakis}}]{Carloni:2015lsa}%
  \BibitemOpen
  \bibfield  {author} {\bibinfo {author} {\bibfnamefont {S.}~\bibnamefont
  {Carloni}}, \bibinfo {author} {\bibfnamefont {F.~S.~N.}\ \bibnamefont
  {Lobo}}, \bibinfo {author} {\bibfnamefont {G.}~\bibnamefont {Otalora}}, \
  and\ \bibinfo {author} {\bibfnamefont {E.~N.}\ \bibnamefont {Saridakis}},\
  }\href {\doibase 10.1103/PhysRevD.93.024034} {\bibfield  {journal} {\bibinfo
  {journal} {Phys. Rev.}\ }\textbf {\bibinfo {volume} {D93}},\ \bibinfo {pages}
  {024034} (\bibinfo {year} {2016})},\ \Eprint
  {http://arxiv.org/abs/1512.06996} {arXiv:1512.06996 [gr-qc]} \BibitemShut
  {NoStop}%
%%CITATION = ARXIV:1512.06996;%%
\bibitem [{\citenamefont {Bahamonde}\ \emph
  {et~al.}(2017{\natexlab{a}})\citenamefont {Bahamonde}, \citenamefont
  {Böhmer}, \citenamefont {Carloni}, \citenamefont {Copeland}, \citenamefont
  {Fang},\ and\ \citenamefont {Tamanini}}]{Bahamonde:2017ize}%
  \BibitemOpen
  \bibfield  {author} {\bibinfo {author} {\bibfnamefont {S.}~\bibnamefont
  {Bahamonde}}, \bibinfo {author} {\bibfnamefont {C.~G.}\ \bibnamefont
  {Böhmer}}, \bibinfo {author} {\bibfnamefont {S.}~\bibnamefont {Carloni}},
  \bibinfo {author} {\bibfnamefont {E.~J.}\ \bibnamefont {Copeland}}, \bibinfo
  {author} {\bibfnamefont {W.}~\bibnamefont {Fang}}, \ and\ \bibinfo {author}
  {\bibfnamefont {N.}~\bibnamefont {Tamanini}},\ }\href@noop {} {\  (\bibinfo
  {year} {2017}{\natexlab{a}})},\ \Eprint {http://arxiv.org/abs/1712.03107}
  {arXiv:1712.03107 [gr-qc]} \BibitemShut {NoStop}%
%%CITATION = ARXIV:1712.03107;%%
\bibitem [{\citenamefont {Bahamonde}\ \emph
  {et~al.}(2018{\natexlab{b}})\citenamefont {Bahamonde}, \citenamefont
  {Marciu},\ and\ \citenamefont {Rudra}}]{Bahamonde:2018miw}%
  \BibitemOpen
  \bibfield  {author} {\bibinfo {author} {\bibfnamefont {S.}~\bibnamefont
  {Bahamonde}}, \bibinfo {author} {\bibfnamefont {M.}~\bibnamefont {Marciu}}, \
  and\ \bibinfo {author} {\bibfnamefont {P.}~\bibnamefont {Rudra}},\
  }\href@noop {} {\  (\bibinfo {year} {2018}{\natexlab{b}})},\ \Eprint
  {http://arxiv.org/abs/1802.09155} {arXiv:1802.09155 [gr-qc]} \BibitemShut
  {NoStop}%
%%CITATION = ARXIV:1802.09155;%%
\bibitem [{\citenamefont {Bahamonde}\ and\ \citenamefont
  {Wright}(2015)}]{Bahamonde:2015hza}%
  \BibitemOpen
  \bibfield  {author} {\bibinfo {author} {\bibfnamefont {S.}~\bibnamefont
  {Bahamonde}}\ and\ \bibinfo {author} {\bibfnamefont {M.}~\bibnamefont
  {Wright}},\ }\href {\doibase 10.1103/PhysRevD.92.084034,
  10.1103/PhysRevD.93.109901} {\bibfield  {journal} {\bibinfo  {journal} {Phys.
  Rev.}\ }\textbf {\bibinfo {volume} {D92}},\ \bibinfo {pages} {084034}
  (\bibinfo {year} {2015})},\ \bibinfo {note} {[Erratum: Phys.
  Rev.D93,no.10,109901(2016)]},\ \Eprint {http://arxiv.org/abs/1508.06580}
  {arXiv:1508.06580 [gr-qc]} \BibitemShut {NoStop}%
%%CITATION = ARXIV:1508.06580;%%
\bibitem [{\citenamefont {Sotiriou}\ \emph {et~al.}(2011)\citenamefont
  {Sotiriou}, \citenamefont {Li},\ and\ \citenamefont
  {Barrow}}]{Sotiriou:2010mv}%
  \BibitemOpen
  \bibfield  {author} {\bibinfo {author} {\bibfnamefont {T.~P.}\ \bibnamefont
  {Sotiriou}}, \bibinfo {author} {\bibfnamefont {B.}~\bibnamefont {Li}}, \ and\
  \bibinfo {author} {\bibfnamefont {J.~D.}\ \bibnamefont {Barrow}},\ }\href
  {\doibase 10.1103/PhysRevD.83.104030} {\bibfield  {journal} {\bibinfo
  {journal} {Phys. Rev.}\ }\textbf {\bibinfo {volume} {D83}},\ \bibinfo {pages}
  {104030} (\bibinfo {year} {2011})},\ \Eprint {http://arxiv.org/abs/1012.4039}
  {arXiv:1012.4039 [gr-qc]} \BibitemShut {NoStop}%
%%CITATION = ARXIV:1012.4039;%%
\bibitem [{\citenamefont {Li}\ \emph {et~al.}(2011{\natexlab{b}})\citenamefont
  {Li}, \citenamefont {Sotiriou},\ and\ \citenamefont {Barrow}}]{Li:2010cg}%
  \BibitemOpen
  \bibfield  {author} {\bibinfo {author} {\bibfnamefont {B.}~\bibnamefont
  {Li}}, \bibinfo {author} {\bibfnamefont {T.~P.}\ \bibnamefont {Sotiriou}}, \
  and\ \bibinfo {author} {\bibfnamefont {J.~D.}\ \bibnamefont {Barrow}},\
  }\href {\doibase 10.1103/PhysRevD.83.064035} {\bibfield  {journal} {\bibinfo
  {journal} {Phys. Rev.}\ }\textbf {\bibinfo {volume} {D83}},\ \bibinfo {pages}
  {064035} (\bibinfo {year} {2011}{\natexlab{b}})},\ \Eprint
  {http://arxiv.org/abs/1010.1041} {arXiv:1010.1041 [gr-qc]} \BibitemShut
  {NoStop}%
%%CITATION = ARXIV:1010.1041;%%
\bibitem [{\citenamefont {Tamanini}\ and\ \citenamefont
  {Boehmer}(2012)}]{Tamanini:2012hg}%
  \BibitemOpen
  \bibfield  {author} {\bibinfo {author} {\bibfnamefont {N.}~\bibnamefont
  {Tamanini}}\ and\ \bibinfo {author} {\bibfnamefont {C.~G.}\ \bibnamefont
  {Boehmer}},\ }\href {\doibase 10.1103/PhysRevD.86.044009} {\bibfield
  {journal} {\bibinfo  {journal} {Phys. Rev.}\ }\textbf {\bibinfo {volume}
  {D86}},\ \bibinfo {pages} {044009} (\bibinfo {year} {2012})},\ \Eprint
  {http://arxiv.org/abs/1204.4593} {arXiv:1204.4593 [gr-qc]} \BibitemShut
  {NoStop}%
%%CITATION = ARXIV:1204.4593;%%
\bibitem [{\citenamefont {Krššák}\ and\ \citenamefont
  {Saridakis}(2016)}]{Krssak:2015oua}%
  \BibitemOpen
  \bibfield  {author} {\bibinfo {author} {\bibfnamefont {M.}~\bibnamefont
  {Krššák}}\ and\ \bibinfo {author} {\bibfnamefont {E.~N.}\ \bibnamefont
  {Saridakis}},\ }\href {\doibase 10.1088/0264-9381/33/11/115009} {\bibfield
  {journal} {\bibinfo  {journal} {Class. Quant. Grav.}\ }\textbf {\bibinfo
  {volume} {33}},\ \bibinfo {pages} {115009} (\bibinfo {year} {2016})},\
  \Eprint {http://arxiv.org/abs/1510.08432} {arXiv:1510.08432 [gr-qc]}
  \BibitemShut {NoStop}%
%%CITATION = ARXIV:1510.08432;%%
\bibitem [{\citenamefont {Harko}\ and\ \citenamefont
  {Lobo}(2010)}]{Harko:2010mv}%
  \BibitemOpen
  \bibfield  {author} {\bibinfo {author} {\bibfnamefont {T.}~\bibnamefont
  {Harko}}\ and\ \bibinfo {author} {\bibfnamefont {F.~S.~N.}\ \bibnamefont
  {Lobo}},\ }\href {\doibase 10.1140/epjc/s10052-010-1467-3} {\bibfield
  {journal} {\bibinfo  {journal} {Eur. Phys. J.}\ }\textbf {\bibinfo {volume}
  {C70}},\ \bibinfo {pages} {373} (\bibinfo {year} {2010})},\ \Eprint
  {http://arxiv.org/abs/1008.4193} {arXiv:1008.4193 [gr-qc]} \BibitemShut
  {NoStop}%
%%CITATION = ARXIV:1008.4193;%%
\bibitem [{\citenamefont {Avelino}\ and\ \citenamefont
  {Sousa}(2018)}]{Avelino:2018qgt}%
  \BibitemOpen
  \bibfield  {author} {\bibinfo {author} {\bibfnamefont {P.~P.}\ \bibnamefont
  {Avelino}}\ and\ \bibinfo {author} {\bibfnamefont {L.}~\bibnamefont
  {Sousa}},\ }\href {\doibase 10.1103/PhysRevD.97.064019} {\bibfield  {journal}
  {\bibinfo  {journal} {Phys. Rev.}\ }\textbf {\bibinfo {volume} {D97}},\
  \bibinfo {pages} {064019} (\bibinfo {year} {2018})},\ \Eprint
  {http://arxiv.org/abs/1802.03961} {arXiv:1802.03961 [gr-qc]} \BibitemShut
  {NoStop}%
%%CITATION = ARXIV:1802.03961;%%
\bibitem [{\citenamefont {Paliathanasis}(2017)}]{Paliathanasis:2017flf}%
  \BibitemOpen
  \bibfield  {author} {\bibinfo {author} {\bibfnamefont {A.}~\bibnamefont
  {Paliathanasis}},\ }\href {\doibase 10.1088/1475-7516/2017/08/027} {\bibfield
   {journal} {\bibinfo  {journal} {JCAP}\ }\textbf {\bibinfo {volume} {1708}},\
  \bibinfo {pages} {027} (\bibinfo {year} {2017})},\ \Eprint
  {http://arxiv.org/abs/1706.02662} {arXiv:1706.02662 [gr-qc]} \BibitemShut
  {NoStop}%
%%CITATION = ARXIV:1706.02662;%%
\bibitem [{\citenamefont {Bahamonde}\ and\ \citenamefont
  {Böhmer}(2016)}]{Bahamonde:2016kba}%
  \BibitemOpen
  \bibfield  {author} {\bibinfo {author} {\bibfnamefont {S.}~\bibnamefont
  {Bahamonde}}\ and\ \bibinfo {author} {\bibfnamefont {C.~G.}\ \bibnamefont
  {Böhmer}},\ }\href {\doibase 10.1140/epjc/s10052-016-4419-8} {\bibfield
  {journal} {\bibinfo  {journal} {Eur. Phys. J.}\ }\textbf {\bibinfo {volume}
  {C76}},\ \bibinfo {pages} {578} (\bibinfo {year} {2016})},\ \Eprint
  {http://arxiv.org/abs/1606.05557} {arXiv:1606.05557 [gr-qc]} \BibitemShut
  {NoStop}%
%%CITATION = ARXIV:1606.05557;%%
\bibitem [{\citenamefont {Bahamonde}\ \emph
  {et~al.}(2017{\natexlab{b}})\citenamefont {Bahamonde}, \citenamefont
  {Böhmer},\ and\ \citenamefont {Krššák}}]{Bahamonde:2017wwk}%
  \BibitemOpen
  \bibfield  {author} {\bibinfo {author} {\bibfnamefont {S.}~\bibnamefont
  {Bahamonde}}, \bibinfo {author} {\bibfnamefont {C.~G.}\ \bibnamefont
  {Böhmer}}, \ and\ \bibinfo {author} {\bibfnamefont {M.}~\bibnamefont
  {Krššák}},\ }\href {\doibase 10.1016/j.physletb.2017.10.026} {\bibfield
  {journal} {\bibinfo  {journal} {Phys. Lett.}\ }\textbf {\bibinfo {volume}
  {B775}},\ \bibinfo {pages} {37} (\bibinfo {year} {2017}{\natexlab{b}})},\
  \Eprint {http://arxiv.org/abs/1706.04920} {arXiv:1706.04920 [gr-qc]}
  \BibitemShut {NoStop}%
%%CITATION = ARXIV:1706.04920;%%
\end{thebibliography}%

\end{document}